\documentclass[
amsmath, amssymb, 
aps, prx,
twocolumn,
superscriptaddress,
longbibliography,
floatfix,
showkeys
]{revtex4-1}

\usepackage{graphicx} 
\usepackage{float}
\usepackage{wrapfig}
\usepackage{array}
\usepackage{diagbox}
\usepackage{chngpage}

\usepackage{braket}
\usepackage{accents}
\usepackage{bm}
\usepackage{mathtools}
\usepackage{upgreek}

\usepackage{color}
\usepackage{hhline}
\usepackage[normalem]{ulem}

\usepackage{csquotes}
\usepackage{datetime}
\usepackage{units}

\newcommand{\Yb}{{}^{171}\textrm{Yb}^+}

\newcommand{\sigx}{\hat{\sigma}_{x}}
\newcommand{\sigy}{\hat{\sigma}_{y}}
\newcommand{\sigz}{\hat{\sigma}_z}
\newcommand{\Id}{{\hat{\mathbb{I}}}}
\newcommand{\sigVect}{{\hat{\boldsymbol{\sigma}}}}
\newcommand{\n}{\textbf{\emph{n}}}

\newcommand{\cleanC}[1]{\hat{C}_{#1}} 
\newcommand{\noisyC}[1]{\tilde{C}_{#1}} 
\newcommand{\errorOp}[1]{\hat{\Lambda}_{#1}} 
\newcommand{\errorOpw}[2]{\hat{\Lambda}^{(#2)}_{#1}} 

\newcommand{\Order}[2][1]{\mathcal{O}\left(#2\right)}
\newcommand{\defeq}{\vcentcolon=}

\newcommand{\abs}[1]{\vert #1 \vert}
\newcommand{\norm}[1]{\| #1 \|}

\newcommand{\rVect}[1]{{\textbf{r}}_{#1}}
\newcommand{\modr}[1]{\norm{{\textbf{r}}_{#1}}}

\newcommand{\V}[1]{{\textbf{V}}_{#1}}
\newcommand{\VNormSq}[1]{\norm{{\textbf{V}}_{#1}}^2}
\newcommand{\R}[1]{{\textbf{R}}_{#1}}
\newcommand{\Ri}[2]{{\textbf{R}}_{#1}^{(#2)}}
\newcommand{\RNormSq}[1]{\norm{{\textbf{R}}_{#1}}^2}
\newcommand{\RiNormSq}[2]{\norm{{\textbf{R}}_{#1}^{(#2)}}^2}

\newcommand{\del}[2]{\delta_{#1}^{(#2)}}
\newcommand{\eps}[2]{\varepsilon_{#1}^{(#2)}}

\newcommand{\eCorr}[1]{{\varepsilon_{C}^{(#1)}}}
\newcommand{\sigCorr}{\sigma_C}
\newcommand{\sigUncorr}{\sigma_U}
\newcommand{\rhoCorr}{\rho_C}
\newcommand{\rhoUncorr}{\rho_U}
\newcommand{\Me}{\mathcal{M}_\varepsilon}
\newcommand{\Mn}{\mathcal{M}_\mathrm{n}}

\newcommand{\stepSum}[1]{\sum^{J}_{#1=1}} 

\newcommand{\E}[1]{\mathbb{E} \left[#1\right]} 
\newcommand{\Var}[1]{\mathbb{V} \left[#1 \right]}
\newcommand{\Cov}[2]{\textrm{Cov} \left(#1,#2 \right)}
\newcommand{\gaussDist}{\sim \mathcal{N}(0,\sigma^2)}
\newcommand{\noiseAve}[1]{\langle #1 \rangle_n}

\newcommand{\exprTwo}[1]{\E{\norm{\rVect{#1}}^2}}
\newcommand{\exprTwoCorr}[1]{\E{\norm{\rVect{C, #1}}^2}}
\newcommand{\exprTwoUncorr}[1]{\E{\norm{\rVect{U, #1}}^2}}
\newcommand{\exprFour}[1]{\E{\norm{\rVect{#1}}^4}}

\newcommand{\exprTwoCorrUncorr}[1]{\E{ \norm{\rVect{U, #1 }}^2  \norm{\rVect{C, #1}}^2  } }

\DeclareRobustCommand{\errorVect}{{\boldsymbol{\epsilon}}}
\DeclareRobustCommand{\nuVect}[1]{{\boldsymbol{\nu}}_{#1}}
\DeclareRobustCommand{\etaVect}{{\boldsymbol{\eta}}}
\DeclareRobustCommand{\etaVectj}[1]{\eta_{#1}}

\let\oldsqrt\sqrt
\def\sqrt{\mathpalette\DHLhksqrt}
\def\DHLhksqrt#1#2{%
\setbox0=\hbox{$#1\oldsqrt{#2\,}$}\dimen0=\ht0
\advance\dimen0-0.2\ht0
\setbox2=\hbox{\vrule height\ht0 depth -\dimen0}%
{\box0\lower0.4pt\box2}}

\begin{document}

\title{Dynamically corrected gates suppress spatio-temporal error correlations as measured by randomized benchmarking}
\author{C. L. Edmunds}
\affiliation{ARC Centre for Engineered Quantum Systems, The University of Sydney, School of Physics, NSW Australia}
\affiliation{Q-CTRL Pty Ltd, Sydney, NSW Australia}
\author{C. Hempel}
\affiliation{ARC Centre for Engineered Quantum Systems, The University of Sydney, School of Physics, NSW Australia}
\author{R. J. Harris}
\affiliation{ARC Centre for Engineered Quantum Systems, The University of Queensland, School of Physics and Mathematics, St Lucia, QLD Australia}
\author{V. M. Frey}
\affiliation{ARC Centre for Engineered Quantum Systems, The University of Sydney, School of Physics, NSW Australia}
\affiliation{Q-CTRL Pty Ltd, Sydney, NSW Australia}
\author{T. M. Stace}
\affiliation{ARC Centre for Engineered Quantum Systems, The University of Queensland, School of Physics and Mathematics, St Lucia, QLD Australia}
\author{M. J. Biercuk}
\thanks{Contact: michael.biercuk@sydney.edu.au }
\affiliation{ARC Centre for Engineered Quantum Systems, The University of Sydney, School of Physics, NSW Australia}
\affiliation{Q-CTRL Pty Ltd, Sydney, NSW Australia}
\date{\today}

\begin{abstract}
Quantum error correction provides a path to large-scale quantum computers, but is built on challenging assumptions about the characteristics of the underlying errors.  In particular, the mathematical assumption of statistically independent errors in quantum logic operations is at odds with realistic environments where error sources may exhibit strong temporal and spatial correlations.  We present experiments using trapped ions to demonstrate that the use of dynamically corrected gates (DCGs), generally considered for the reduction of error magnitudes, can also suppress error correlations in space and time throughout quantum circuits. We present a first-principles analysis of the manifestation of error correlations in randomized benchmarking, and validate this model through experiments performed using engineered errors.  We find that standard DCGs can reduce error correlations by $\sim50\times$, while increasing the magnitude of uncorrelated errors by a factor scaling linearly with the extended DCG duration compared to a primitive gate. We then demonstrate that the correlation characteristics of intrinsic errors in our system are modified by use of DCGs, consistent with a picture in which DCGs whiten the effective error spectrum induced by external noise. 
\end{abstract}

\maketitle

\section{Introduction}

Suppressing and correcting errors in quantum circuits is a critical challenge driving a substantial fraction of research in the quantum information science community. These efforts build on quantum error correction (QEC) and the theory of fault tolerance \cite{Shor:1995, Shor:1996, Steane:1996, Steane:1996b, Aharonov:1997, Gottesman:1998} as the fundamental developments that support the concept of large-scale quantum computation~\cite{Preskill:1998,
QECLidar2013, Campbell:2017}. In combination, these theoretical constructs suggest that so long as the probability of error in each physical quantum information carrier can be reduced below a threshold value, a properly executed QEC protocol can detect and suppress logical errors to arbitrarily low levels, and hence enable arbitrarily large computations. Underlying this proposition is an assumption that errors are statistically independent, i.e., the emergence of a qubit error at a specific time is uncorrelated with errors arising in other qubits or at any other time in the computation. Error correlations that decay with distance between qubits (spatially) can induce simultaneous multi-qubit errors~\cite{Preskill_correlations2013}, and correlations that decay with circuit length (temporally) have been shown to produce more rapid accumulation of net circuit errors~\cite{Wallman:2015, Proctor:2017}. 

The practicality of the assumption of uncorrelated errors has long been questioned, as laboratory sources of noise commonly exhibit strong temporal correlations, captured through spectral measures exhibiting high weight at low frequencies. As such, coherent errors induced by low frequency noise and miscalibrations have recently become a larger focus of research, with their detrimental effects on QEC implementations being examined ~\cite{Wallman:2015, Greenbaum:2018,Huang:2018,Chubb:2018} and first ideas targeting their suppression emerging \cite{Debroy:2018, Majumder:2019}. Attempts to address these errors in the theory of quantum error correction are challenging and results to date suggest that revision of postulated fault-tolerant thresholds may be required~\cite{Preskill2009, Preskill2006} relative to more optimistic predictions that have recently emerged~\cite{Fowler_Surface}. Indeed, when implicit assumptions that errors are both spatially and temporally uncorrelated are weakened, the value of a tolerable error threshold can change from some value $\varepsilon$ to $\varepsilon^2$, easily leading to order-of-magnitude decreases in the acceptable error rates~\cite{Preskill:1998}.

The adverse effect of correlated errors on error correction procedures has been observed in the context of a repetition code both experimentally~\cite{Schindler:2011} -- where they were seen to effectively negate any advantage obtained from iterative error correction -- and theoretically~\cite{Greenbaum:2018}, where an increase in the logical failure rate was identified. Furthermore, while a recent full-scale numerical simulation has shown that coherent errors at the physical layer can, in fact, be overcome by topological error correcting codes~\cite{Bravyi:2018}, large numbers of physical qubits are required with error rates that are \emph{uniformly} sub-threshold.  The emerging message is that, while correlated errors do not invalidate the use of QEC, their presence can significantly increase the requisite overhead, and may reduce the tolerable magnitude of physical qubit errors.

In this manuscript, we demonstrate experimentally that using a low-level abstraction known as a dynamically corrected gate (DCG), we can suppress error correlations in addition to error magnitudes. Replacing \enquote{primitive} physical quantum gate operations with logically equivalent DCGs~\cite{Brown2004, Khodjasteh2009dcg, True, DasSarmaGate, SoareNatPhys2014} forms a ``virtual'' layer wherein error characteristics can be modified (``virtualized'') before the application of QEC~\cite{Preskill_Layered, JonesPRX2012}. We present a novel first-principles analysis of Clifford randomized benchmarking~\cite{Emerson2005, Dankert:2009} in order to quantitatively model the impact of error correlations on simple experimental observables, building on concepts in~\cite{Ball:2016}.  Specifically, we identify that error correlations are manifested in the scaling of the distribution over sequence randomizations, at fixed sequence length, with measurement averaging. We validate this framework using randomized benchmarking experiments performed with a single trapped Ytterbium ion.  We then demonstrate that the replacement of the individual Clifford operations within each sequence with logically equivalent DCGs modifies the error correlation signatures such that they are experimentally consistent with the presence of uncorrelated errors. Single-qubit experiments performed under engineered noise with tunable correlation characteristics show consistent reduction in the correlated error component when switching from primitive to DCG sequences. We explain this behaviour using a framework that describes the action of DCGs at the operator level~\cite{Kabytayev2014,SoareNatPhys2014,ViolaFFF} as whitening the effective error spectrum experienced by each gate. Finally, we demonstrate that using DCGs in sequence construction reduces spatial error correlations between qubits, through simultaneous randomized benchmarking on five trapped ion qubits. These results provide direct and strong evidence that the use of dynamically protected physical qubit operations in a layered architecture for quantum computing~\cite{JonesPRX2012} can facilitate the successful application of existing QEC theory with only minimal revision on the path to fault-tolerant quantum computation.

\section{Identifying signatures of error correlations in circuits}
\label{sec:correlations_theory}

\begin{figure*}[t!]
	\centering
	\includegraphics[scale=1]{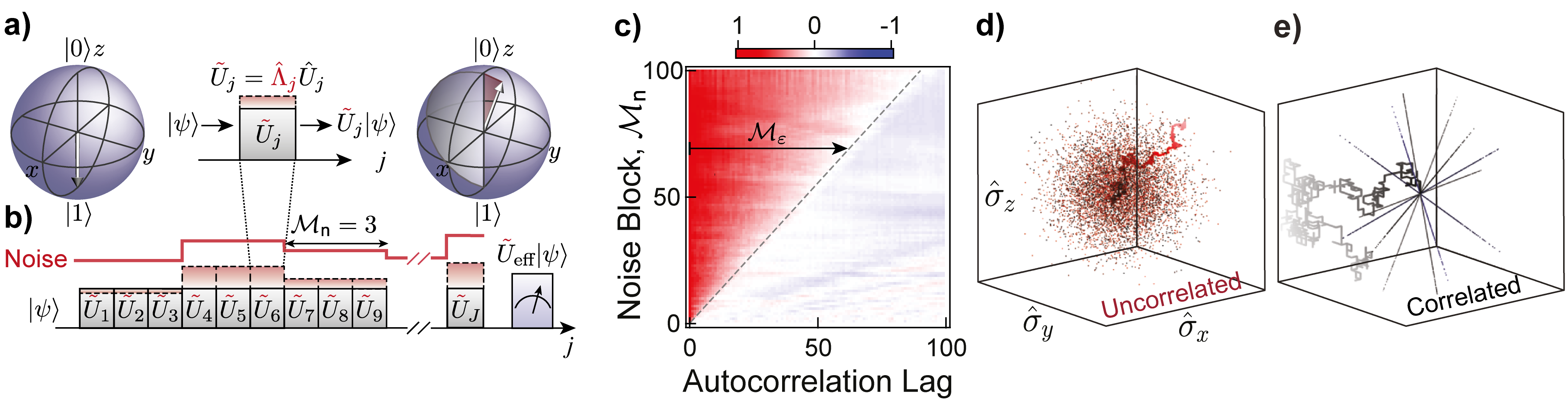}
	\caption{\textbf{Translation of noise correlations to error correlations in quantum circuits.}
\textbf{a,} A single operation applied to a qubit in the presence of noise $\tilde{U}_j$ can be decomposed into an error operator $\errorOp{j}$ and the target operation $\hat{U}_{j}$. Bloch spheres schematically illustrate the effect of an imperfect $\pi$-rotation about the $x$-axis acting on input state $\ket{1}$, with dark shading indicating an over-rotation error. \textbf{b,} Noise (red line) exhibiting non-zero temporal correlation of length $\Mn=3$, quantized in units of gate operations, acts on a quantum circuit composed of sequentially applied unitary operations. The resultant errors accumulate and lead to a noisy effective operator $\tilde{U}_\mathrm{eff}$, whose effect is determined through a projective measurement at the end of the circuit. 
\textbf{c,} Translation of correlations in a noise process to correlations in the magnitude of the circuit error vector, $\norm{\errorVect_j}$. The error vector for each gate of a randomly composed sequence of 1000 primitive gates under a noise process with noise correlation length $\Mn$ is calculated and the autocorrelation function of the magnitude of the error vector, $\E{\norm{\errorVect_{j_1}}\norm{\errorVect_{j_2}}}$, is shown for the first 100 gates.
\textbf{d, e,} Random walks for the extreme error correlation cases, \textbf{d,} $\Me = 1$ (uncorrelated) and \textbf{e,} $\Me = J$ (fully correlated).
Final walk displacements of eight sequences, each with 1000 error realizations, are shown along with the full walk for a single sequence that is common between the two cases.
}
	\label{fig:noise2errorsRW}
\end{figure*}

We begin by laying out the challenge of establishing clear quantitative metrics allowing the identification of error correlations in quantum circuits.  As a first step we analyze how correlations in a physical noise process translate to correlations in the resultant unitary errors within a circuit of $j=1,\ldots,J$ gates. In our model, any noisy operation $\tilde{U}_j$ within the circuit can be decomposed into the ideal operator $\hat{U}_j$ and an error operator $\errorOp{j}$, such that \mbox{$\tilde{U}_j = \errorOp{j}\hat{U}_j$}. Here, \mbox{$\hat{U}_j \equiv \hat{U}(\n_j, \theta_j)$} rotates the state vector by angle $\theta_j$ around an arbitrary axis $\n_j$ on the Bloch sphere. Considering unitary semiclassical noise processes, the error component in each operation can be written as 
\mbox{$\hat{\Lambda}_{j}=\exp\left\{{i\sum_{\alpha=1}^{\infty}[\errorVect_j]_{\alpha}\cdot \sigVect}\right\}$}, 
with $\sigVect$ the vector of Pauli matrices, $\alpha$ an index denoting the Magnus expansion order~\cite{GreenNJP2013}, and $\errorVect_j$ the error vector characterizing the strength and nature (affected quadrature) of the error~\cite{Kofman2004, GreenPRL2012, GreenNJP2013, ViolaFFF}. A quantum circuit experiences temporally correlated errors if the values of $\errorVect_j$ across the circuit (in space or time) exhibit non-zero correlations.

Our approach to measuring error correlations is built on common quantum verification protocols employed to infer the average behavior of gate operations~\cite{Emerson2005, Knill2008, MagesanInterleaved, Emerson2011, Magesan:2012, Merkel:2013, Kimmel:2014, Wallman:2015, Sheldon:2016, BlumeKohout:2017, Mavadia:2017, Onorati:2019}. Restricting our analysis to the single-qubit case, error correlations between gates may occur in these protocols  when physical noise processes exhibit strong correlations in time.  We demonstrate this numerically by calculating the error vector $\errorVect_j$ for each operation in a single-qubit randomized benchmarking sequence exposed to detuning ($\sigz$) noise with a variable block-correlation length, $\Mn$; this is defined to be the number of gates over which the noise strength is constant within the sequence. The sequence is assembled from the 24 Clifford operations comprising combinations of $\pi$ and $\pi/2$-rotations about the $x,y$ and $z$-axes of the Bloch sphere, and an identity gate $\Id$.  Calculating the autocorrelation function of the error vector's magnitude throughout a sequence reveals strong correlations over a length of gates, $\Me$, which appear to scale linearly with the correlation length of the input noise process, $\Mn$ (Fig.~\ref{fig:noise2errorsRW}c). This behavior suggests a linear mapping from noise correlations to error correlations in conventional settings. As a prelude to future demonstrations in this manuscript, we note that if the individual Clifford gates are replaced by DCGs, this simple linear mapping from input noise correlations to output error correlations breaks down. 

In general, the primary limitation one faces in accessing information about $\Me$ in a physical experiment is that using standard, projective measurements at the end of a circuit will limit the ability to probe correlations that arise throughout the circuit's execution.  Most experimental quantum verification routines suffer from exactly this limitation, and primarily measure the average difference between a qubit state transformed under an imperfect operation and a predetermined target state at the end of the protocol (Fig.~\ref{fig:noise2errorsRW}a). However, as we will illustrate in the following, there is additional useful information present in the outcomes of randomized benchmarking measurement routines that may be employed to extract novel insights about error correlations appearing during the sequence. 

The key underlying concept is that in a randomized benchmarking sequence built up from many operations, the resultant net state transformation in the presence of noise, $\tilde{U}_\mathrm{eff}\ket{\psi}$ (Fig.~\ref{fig:noise2errorsRW}b), is determined by an interplay of both the sensitivity of each individual operation to the noise~\cite{GreenNJP2013} and the impact of the sequence structure on error accumulation~\cite{Ball:2016, Wallman:2016, Mavadia:2017}. Specifically, nominally equivalent randomized benchmarking sequences (constructed to perform the same net operation) exhibit variations in correlated-noise susceptibility that are analytically calculable and verifiable in experiments. We use this variability and the behavior under experimental averaging to extract a signature of error correlations within quantum sequences. 

\subsection{Random walk formalism for error accumulation}

We present a first-principles analysis to directly link measurement outcomes for single-qubit randomized benchmarking sequences to the nature of the underlying error correlations quantified by $\Me$, expanding the formalism introduced in reference~\cite{Ball:2016}. We consider randomized benchmarking sequences composed of $J$ single-qubit Clifford operations, \mbox{$\prod_{j=1}^J \cleanC{\etaVectj{j}} = \Id$}, with the vector $\etaVect$ containing labels for the 24 Clifford operations, \mbox{$\etaVectj{j} \in \{1, 2, \dots, 24\}$}. A final gate is pre-calculated to yield a net identity operation for the sequence, such that in the absence of error the final qubit state will be the same as the initial state. Due to imperfections in the operations, the physically implemented gates $\noisyC{\etaVectj{j}}$ differ from the ideal gates by an error map \mbox{$\noisyC{\etaVectj{j}}= \errorOp{j} \cleanC{\etaVectj{j}}$}. 

The accumulation of errors throughout a sequence can be represented by a sequence-dependent ``random walk'' in three-dimensional Pauli-error space; the net walk length can then be related to the final sequence error~\cite{Ball:2016}. For a particular realization of the error $i$, this walk is captured by the vector 

\begin{equation}
    \label{eq:random_walk}
    \Ri{\textrm{3D}}{i} = \stepSum{j} \eps{j}{i} \rVect{\textrm{3D}, j}
\end{equation}
with gate error values \mbox{$\eps{j}{i}\gaussDist$} sampled from a zero-mean Gaussian distribution with rms value $\sigma$. It will be shown in Section~\ref{sec:2D_walks} that this leads to an average, randomized benchmarking error per gate $\propto\sigma^2$. 
Here, the values of $\rVect{\textrm{3D}, j}$ are unit-length vectors that define the sequence-specific random walk steps; they can be calculated deterministically for any randomized benchmarking sequence, irrespective of the strength or correlation characteristics of the gate errors.
In a circumstance where the normalized error takes a consistent value $\eps{j}{i}\equiv 1$, the length of the $J$-step walk created by these steps is an intrinsic property of the sequence and will be shown to act as a proxy for its susceptibility to correlated errors. Examining individual randomized benchmarking sequences reveals the idiosyncratic nature of their walks; certain randomizations exhibit long walks, while others have walks that terminate near the origin, solely determined by the structure of the sequence and the form of the error channel. Accordingly, in the presence of correlated errors we expect a wide variance of outcomes, determined by the underlying structures of the randomly selected sequences. The general framework linking this Pauli walk to accumulated error was experimentally validated in~\cite{Mavadia:2017}.

\subsection{Signatures of error correlations}
We identify that the key measurable signature of error correlations arises in the process of experimental averaging over repetitions of a sequence, and hence over different realizations of the error. In order to understand this, we begin by examining how error correlations impact the random walk introduced above, and how the behavior of that walk changes with experimental averaging.

Gate errors induce the mapping $\rVect{\textrm{3D}, j} \to \Ri{\textrm{3D}}{i}$; the term $\eps{j}{i}$ in Eq.~\eqref{eq:random_walk} can change the direction and scale the magnitude of each step in the random walk. Thus correlations in $\eps{j}{i}$ are translated into correlated modifications of the steps in $\Ri{\textrm{3D}}{i}$.  To see the effect of correlations in the error process, we calculate the locus of walk termination points for eight different sequences and 1000 error realizations, shown in \mbox{Fig.~\ref{fig:noise2errorsRW}d,e}. In the presence of errors whose magnitudes are constant across all gates in a given benchmarking sequence, the error \mbox{$\eps{j}{i}\equiv\eps{}{i}$} rescales all steps in the walk uniformly, such that all termination points for a given sequence fall on a line (Fig.~\ref{fig:noise2errorsRW}e). The walk terminations for the same sequence are thus dominated by the underlying sequence structure (``rays'' in Fig.~\ref{fig:noise2errorsRW}e). By contrast, in the presence of uncorrelated errors where $\eps{j}{i}$ changes randomly for each step, the termination points appear randomly distributed in Pauli space for different realizations of the error (Fig.~\ref{fig:noise2errorsRW}d).

These differences will manifest in an experiment that averages the experimental performance of a set of sequences over many different realizations of an error-inducing noise process. In the case of correlated errors, the preservation of sequence-structure dependence in the sequence error leads to a broad distribution of outcomes over different randomized benchmarking sequences.  This breadth is maintained even when averaging experiments together over various realizations of the random but temporally correlated errors. In contrast, for uncorrelated errors, the random, formless distribution of walk termination points over the same set of sequences implies that averaging over experiments would result in a spread of outcomes that grows narrower as the experiment number increases, consistent with the central limit theorem.  It is therefore in the distribution over measured results of noise-averaged, randomized benchmarking sequences that the signatures of error correlations between gates within a sequence will appear.  In Sections~\ref{sec:2D_walks} and~\ref{sec:theory_to_laboratory} we will describe how this phenomenology can be accessed through a modified analysis of conventional randomized benchmarking experiments.

\subsection{Mapping to measurable quantities}
\label{sec:2D_walks}
We now link the random-walk framework to measurements commonly performed in the laboratory -- a single projective measurement in the qubit basis. Such measurements are unaffected by rotations about the $z$-axis, i.e., they are phase invariant. Consequently, this type of projective measurement is insensitive to the component of the random walk oriented along the $\sigz$-axis, and instead probes a two-dimensional projection of the walk onto the $\sigx\sigy$-plane of Pauli-error space~\cite{Mavadia:2017}. Considering a measurement routine involving averaging a single sequence over $n$ realizations of the error, we may relate the two-dimensional walk length to the projective measurement results as,
\begin{equation}
	\label{eq:state_fid}
	\mathcal{P} = 1 - \noiseAve{\RNormSq{\textrm{2D}}} + \Order{\sigma^4},
\end{equation}
where \mbox{$\noiseAve{\cdot}$} is an average over $n$ instances of the error process, \mbox{$\mathcal{P} \defeq 1 - \noiseAve{P(\ket{1})}$} is the measurable, noise-averaged sequence ``survival probability'' when the qubit is initialized in the state $\ket{0}$, $\sigma$ is the rms of the normally distributed errors, and $\R{\textrm{2D}}$ denotes the random walk in the $\sigx\sigy$-plane of Pauli-error space. For simplicity, we will proceed by referring to  $\R{\textrm{2D}}$, and its individual steps $\rVect{\textrm{2D}, j}$, simply as $\R{}$ and $\rVect{j}$ respectively.

We analyze in detail three distinct error correlation regimes for a unitary error channel with values \mbox{$\eps{j}{i}\gaussDist$}: (i) $\Me=J$, identically correlated errors with fixed, constant magnitude over a sequence and rms value $\sigCorr$; (ii) $\Me=1$, uncorrelated, normally distributed errors that change randomly between each gate in a sequence with rms value $\sigUncorr$; and (iii) statistically independent, contemporaneous correlated and uncorrelated error processes such that the relative strengths $\sigCorr$ and $\sigUncorr$ determine the effective error correlation length.

The expression for survival probability in Eq.~\eqref{eq:state_fid} can be used to calculate the distribution of survival probabilities without modification for both regime (i) and (ii) simply by using the appropriately calculated random walks. In the limit of long sequences and many noise averages (large $J$ and $n$), the noise-averaged survival probability is Gamma distributed over different, nominally equivalent, sequence randomizations~\cite{Mavadia:2017}; the shape and scale parameters of the distribution, $a$ and $b$ respectively, can be calculated from first principles using the particulars of the sequence, noise averaging, and error characteristics.  For these two limiting cases of identically correlated errors over a sequence and uncorrelated errors changing randomly between gates, the respective survival probabilities are sampled from Gamma distributions shaped according to
\begin{subequations}
\label{eq:gamma_dists}
\begin{align}
	\mathcal{P}_{C} &\sim \Gamma(a = 1, b = \tfrac{2}{3}J\sigma^2),   \\
	\mathcal{P}_{U} &\sim \Gamma(a = n, b = \tfrac{2}{3n}J\sigma^2).
\end{align}
\end{subequations}	
From these expressions, the variance and expectation values of the distribution over sequence randomizations can be calculated. To leading order, both distributions exhibit the same mean value \mbox{$\mathbb{E} = ab$}, giving a randomized benchmarking average gate error of $\frac{2}{3}\sigma^2$. However, the distributions diverge in the second moment \mbox{$\mathbb{V} = ab^2$}.

We may now derive the properties of the distribution associated with regime (iii) by considering two independent walks; one is induced by the correlated error component $\Ri{C}{i}$, and the other by the uncorrelated component $\Ri{U}{i}$. To begin, it is convenient to note that in the case of a correlated, fixed error process over a sequence, it is possible to factor out the constant error strength from the random walk for a particular realization of the error~\cite{Ball:2016},
\begin{equation}
    \Ri{C}{i} =  \eps{C}{i} \stepSum{j} \rVect{j} = \eps{C}{i} \V{}.
\end{equation}
We thus introduce $\V{}$ to describe the sequence-specific walk, defined by the steps $\rVect{j}$ that remain invariant under different realizations of the error process (Fig.~\ref{fig:noise2errorsRW}e).  This separability is not achievable in the presence of uncorrelated errors due to the randomization of each step in the walk by the error process.  The expression for survival probability can then be expanded in terms of these independent walks to second order in $\sigCorr,\,\sigUncorr$ as
\begin{align}
    \label{eq:dual_error_fid}
    \mathcal{P} &= 1 - \noiseAve{ \norm{\Ri{U}{i} + \eCorr{i} \V{}}^2  }   \nonumber\\
    &= 1 - \noiseAve{ \RiNormSq{U}{i}} - \sigCorr^2  \VNormSq{},
\end{align} 
where the cross-term is identically zero using \mbox{$\noiseAve{\eCorr{i}} = 0$}.

For all three correlation regimes, higher-order terms and cross-terms contribute to the second moment of the distribution and have been calculated analytically (Table~\ref{tab:statistical_moments}). These terms reduce to those calculated using the Gamma distributions in Eq.~\ref{eq:gamma_dists} in the limit of large $J$ and $n$, with $J\gg n$.  On inspection, we expect that in the presence of uncorrelated errors the variance will narrow with increasing $n$, while it will remain fixed in the presence of correlated errors.  Such differences in scaling of a variance measure with averaging are reminiscent of the manifestation of noise correlations in other physical quantities, e.g., the Allan variance used in precision frequency metrology~\cite{Allan:1966, Rutman1978}.  Our analysis therefore highlights that calculating the variance of measurements of randomized benchmarking survival probabilities for different sequences, and exploring how this variance changes with experimental averaging, can give insights into the underlying error correlations.  The functional dependence of the distribution variance with $n$ will be employed throughout the remainder of this work as a key signature of error correlations in standard randomized benchmarking.

\begin{table}[!t]
\center
    \begin{tabular}{|l||c|c|}
 	\hline
	\parbox[c][30pt][c]{\textwidth/6}{\textbf{Error Type}}
	& $1 - \E{\mathcal{P} }$
	& $\Var{\mathcal{P} }$
	\\ \hhline{|=||=|=|}
	\parbox[c][30pt][c]{\textwidth/6}{(i) Fully Correlated,\\$\Me=J$}
	& $ \frac{2}{3} J \sigCorr^2 $
	& $ \frac{2}{9} \frac{(n+2)}{n} J(2J - 1)  \sigCorr^4 $
	\\ \hline
	\parbox[c][30pt][c]{\textwidth/6}{(ii) Uncorrelated,\\$\Me=1$}
	& $ \frac{2}{3} J \sigUncorr^2 $
	& $ \frac{2}{9n} J \left(4 + 2J + n \right) \sigUncorr^4 $
	\\ \hline
	\parbox[c][40pt][c]{\textwidth/6}{(iii) Correlated +\\Uncorrelated}
	&$ \frac{2}{3} J (\sigUncorr^2 + \sigCorr^2 )  $
	&	\parbox[c][40pt][c]{\textwidth/7}{$ \Var{ \mathcal{P}_U} + \Var{  \mathcal{P}_C }$\\ \parbox[c][20pt][c]{\textwidth/7}{$+ \frac{4}{9} J \sigCorr^2 \sigUncorr^2$}}
	\\ \hline				
	\end{tabular}
\caption{The statistical moments for the distribution of noise-averaged sequence survival probabilities with different error correlation lengths - fully correlated across the sequence, completely uncorrelated values between gates, and a combination of two independent error processes in the same quadrature, one correlated and one uncorrelated. The variance for case (iii) incorporates contributions from each error source individually, $\Var{ \mathcal{P}_C}, \Var{ \mathcal{P}_U}$, as well as a cross-term.}
\label{tab:statistical_moments}
\end{table}

In the next section we demonstrate how the model can be updated to connect to realistic laboratory noise models.

\subsection{Modelling realistic laboratory error models}
\label{sec:theory_to_laboratory}
Building on the general framework introduced above, we introduce new first-principles calculations connecting the theoretical model for gate \emph{error} with actual, error-inducing \emph{noise} in experiments. We determine the sequence walk in the presence of arbitrary, unitary error maps,  incorporating the possibility of multi-axis and gate-dependent errors. This facilitates the analysis of experimental measurements performed subject to the most common noise sources encountered in the laboratory.

We consider two physically motivated noise processes that can occur throughout a randomized benchmarking sequence.  First, frequency detuning noise -- either on the qubit's resonant frequency or the frequency of the control field used to drive qubit gate operations -- creates an off-resonance error between the qubit and control. Second, amplitude noise, which may arise from coupling-strength variations or drifts and miscalibrations in the control, results in an over- or under-rotation error of the qubit state vector. Both of these represent ``concurrent'' noise sources (i.e., applied simultaneously with the execution of a gate), which ultimately produce complex gate-dependent errors.

In general, depending on their underlying cause, both frequency detuning and amplitude noise processes may possess temporally correlated and uncorrelated components. Correlated noise sources include miscalibrations, magnetic field drifts, and temperature drifts in control systems, while uncorrelated noise often stems from electrical noise or local environmental sources, e.g., anomalous heating in ion traps \cite{Sedlacek:2018} or two-level system (TLS) fluctuators in superconducting qubits \cite{Schloer:2019, Burnett:2019}. 

To now examine the impact of these physical noise processes on the behavior of the sequence survival-probability distributions, we proceed by explicitly calculating the translation between the physical noise strength, $\del{j}{i}\sim\mathcal{N}(0, \rho^2)$, and the effective sequence errors at the core of our model $\varepsilon = \varepsilon(\delta)$. In our notation, $\rho$ is used to denote the rms magnitude of the noise, distinguishing it from the rms magnitude of the error operator $\sigma$. Our calculations incorporate the fact that single-axis noise (e.g., detuning) present during a non-commuting operation generally results in a multi-axis error process. Furthermore, physical implementations of Clifford operations typically employ variable gate durations, resulting in gate-dependent error operators.

\begin{table*}[!t]
\center
    \begin{tabular}{|c||c|c|}
 	\hline
	\parbox[c][30pt][c]{\textwidth/6}{\textbf{Error Type} }
	& \textbf{$\rho\to\sigma$ Translation for} $\mathbb{E}$
	&  \textbf{$\rho\to\sigma$ Translation for} $\mathbb{V}$
	\\ \hhline{|=||=|=|}
	\parbox[c][30pt][c]{\textwidth/6}{(i) Fully Correlated, \\ $\Mn=J$}
	& $  \sigCorr^2 =  \frac{3}{2} \exprTwo{j} \rhoCorr^2 $
	& $  \sigCorr^4 =  \frac{9}{2} \frac{ \exprFour{j} + (J - 2) \exprTwo{j}^2}{2J-1} \rhoCorr^4$
	\\ \hline
	\parbox[c][30pt][c]{\textwidth/6}{(ii) Uncorrelated, \\ $\Mn=1$}
	& $  \sigUncorr^2 = \frac{3}{2} \exprTwo{j} \rhoUncorr^2 $
	& $  \sigUncorr^4 =  \frac{9}{2} \frac{ (2+n)\exprFour{j} + (J-1-n) \exprTwo{j}^2  }{4+2J+n} \rhoUncorr^4 $
	\\ \hline
	\parbox[c][30pt][c]{\textwidth/6}{(iii) Correlated + \\ Uncorrelated}
	&$ (\sigCorr^2 + \sigUncorr^2) = \frac{3}{2} \left( \E{\norm{\rVect{U, j}}^2} \rhoUncorr^2 + \E{\norm{\rVect{C, j}}^2} \rhoCorr^2 \right) $
	&$ \sigCorr^2 \sigUncorr^2 = \frac{9}{2} \Cov{ \norm{\rVect{U, j}}^2 }{\norm{\rVect{C, j}}^2 } \rhoCorr^2 \rhoUncorr^2$
	\\ \hline				
	\end{tabular}
\caption{The translation from the rms value of a physical noise process, $\rho$, with correlation length $\Mn$, to the rms value of the gate error, $\sigma$, used to calculate the first and second moments of noise-averaged sequence survival probabilities. The values $\rhoCorr, \rhoUncorr$ represent the rms magnitudes of the correlated and uncorrelated noise processes respectively. Similarly, the terms $\rVect{U, j}, \rVect{C, j}$ represent the random walk steps for the different noise processes.  Full details of the derivation of the relevant random walk step expectation values, $\exprTwo{j}, \, \exprFour{j},$ and \mbox{$\Cov{ \norm{\rVect{U, j}}^2 }{\norm{\rVect{C, j}}^2 }$} for the specific noise models employed in our verification experiments are presented in Appendix~\ref{app:expr_calculation}.}
\label{tab:noise_to_error}
\end{table*}

In this setting, the error $\eps{j}{i}$ employed in Eq.~\eqref{eq:random_walk} is replaced by the physical noise strength $\del{j}{i}$.  As a result, the previously unit-length steps $\rVect{\textrm{3D}, j}$ now take more complex, but still analytically calculable, values due to the gate-dependence and multi-axis character of the errors induced by concurrent noise processes.  For a particular noise process we calculate the associated random walk, which enables a mapping of the rms magnitude of the physical noise $\rho$ to an updated rms value of the error $\sigma$. Appendix~\ref{app:revised_RB_theory} describes the formalism to calculate the noise-to-error translation in standard Clifford gates for an arbitrary, unitary error process. Table~\ref{tab:noise_to_error} summarizes the results which, when combined with the expressions from Table~\ref{tab:statistical_moments}, can be used to predict both the expectation and the variance of the distribution of survival probabilities over sequence randomization.  

\section{Experimental Implementation}
\label{sec:experimental_setup}
\subsection{Randomized benchmarking on $\Yb$ qubits}
We perform experiments using a qubit encoded in the $^2\mathrm{S}_{1/2}$ hyperfine ground states of a single laser-cooled $\Yb$ ion confined in a linear Paul trap, with the computational basis states defined as \mbox{$\ket{0} \defeq \ket{F=0, m_F = 0}$} and \mbox{$\ket{1} \defeq  \ket{F=1, m_F = 0}$}.
Laser cooling, state initialization to $\ket{0}$, and detection are performed using a laser at 369\,nm that couples the ${}^{2}{\textrm{S}}_{1/2} \ket{{F}=1}$ ground state to the first excited state ${}^2\mathrm{P}_{1/2} \ket{ {F}=0}$. As the ion selectively fluoresces when it is projected to the upper, ``bright'' qubit state $\ket{1}$, one can distinguish between the two basis states by counting the number of emitted photons during the detection period.
Single-ion qubit state detection is performed in a time-resolved manner~\cite{Wolk:2015, Mavadia:2017} using an avalanche photodiode; multi-ion data employs an EMCCD camera and processing through a Random Forest classifier from the scikit-learn framework~\cite{scikit-learn}.

Qubit rotations are driven via a microwave field near 12.6\,GHz generated by a Vector Signal Generator (VSG). Using an internal baseband generator, we program arbitrary rotations of the qubit via \textit{IQ} modulation. Rotations about the $z$-axis are implemented as instantaneous, pre-calculated \textit{IQ} frame shifts. Randomized benchmarking sequences composed from Clifford operations are pre-loaded into the VSG and mapped to the desired physical operations prior to the recording of each data set. The experiments in this manuscript are performed using $k$ sequences each comprising $J$ operations. The first $J-1$ gates are randomly composed Clifford operations, $\cleanC{\etaVectj{j}}$, and the final operation, \mbox{$\cleanC{\etaVectj{J}} =  (\prod_{j=1}^{J-1}\cleanC{\etaVectj{j}})^{\dagger}$}, is selected such that the sequence implements the identity in the absence of error. A full list of the Clifford operations and their physical implementations can be found in the \textit{Supplementary Materials} of reference \cite{Ball:2016}.  Typical, single-qubit randomized benchmarking experiments with primitive gates  achieve a baseline result of $p_{\textrm{RB}}\approx 1.9\times 10^{-5}$ in our system (Appendix~\ref{app:single_qubit_RB}).

\subsection{Verifying error correlation signatures with engineered errors}\label{sec:histograms}

The key signature of the presence of temporally correlated errors appears in the variance of the distribution over sequence survival probabilities and its scaling with experimental averaging; averaging reduces the variance in the case of uncorrelated errors, but has limited impact when errors exhibit strong temporal correlations. 

We begin our experimental study by engineering experimental noise sources to test and verify the predictions of the theoretical model presented in Sec.~\ref{sec:correlations_theory}. We perform standard randomized benchmarking, but engineer detuning and control-amplitude noise with different user-defined bandwidths. All noise values are generated numerically, and are sampled from a zero-mean Gaussian distribution $\mathcal{N}(0,\rho^2)$ with rms strength $\rho$. Off-resonance errors are induced via fractional detuning noise present during the application of the randomized benchmarking sequence, $\delta = (\Delta/\Omega)$, set by the frequency detuning $\Delta$ between the qubit transition and the microwave source in units of the Rabi frequency, $\Omega$. Over-rotation errors are produced by amplitude noise in the microwave control field, effectively changing $\Omega$. Two limiting noise bandwidths are treated: maximally correlated noise, $\Mn=J$, and uncorrelated noise, $\Mn \leq 1$. For the detuning (control-amplitude) noise process, the correlated noise component is engineered using a constant offset in the VSG microwave frequency (amplitude) over the entire sequence, and the uncorrelated noise is applied via an external FM (AM) modulation input, and changes value every primitive $\pi/2$-time. The relevant random walk steps calculated for these noise processes and used in modelling our experimental measurements are found in Table~\ref{tab:expected_error_steps} of Appendix~\ref{app:revised_RB_theory}.

Instead of simply calculating the randomized benchmarking decay rate, $p_\textrm{RB}$ derived from fitting to the mean of the distribution over different values of $J$, we instead focus on analyzing our data to extract information that is otherwise generally discarded in averaging processes. In each individual measurement, the qubit is initialized in state $\ket{0}$ via optical pumping and one of $k=50$ randomized benchmarking sequences with $J=100$ gates is applied in the presence of engineered noise. A final projective measurement in each experiment yields a discretized qubit state measurement, which is used to infer the probability of finding the qubit in state $\ket{1}$ by repeating the experiment $r=220$ times under application of the same engineered noise realization (reducing quantum projection noise). The survival-probability measurement outcomes for each sequence are then averaged over a variable number up to $n=200$ different realizations of noise possessing the same engineered correlations.  This process is repeated for all $k=50$ sequences, allowing us to calculate the distribution variance $\mathbb{V}_{k}^{(n)}$.

\begin{figure}[t]
	\centering
	\includegraphics[scale=1]{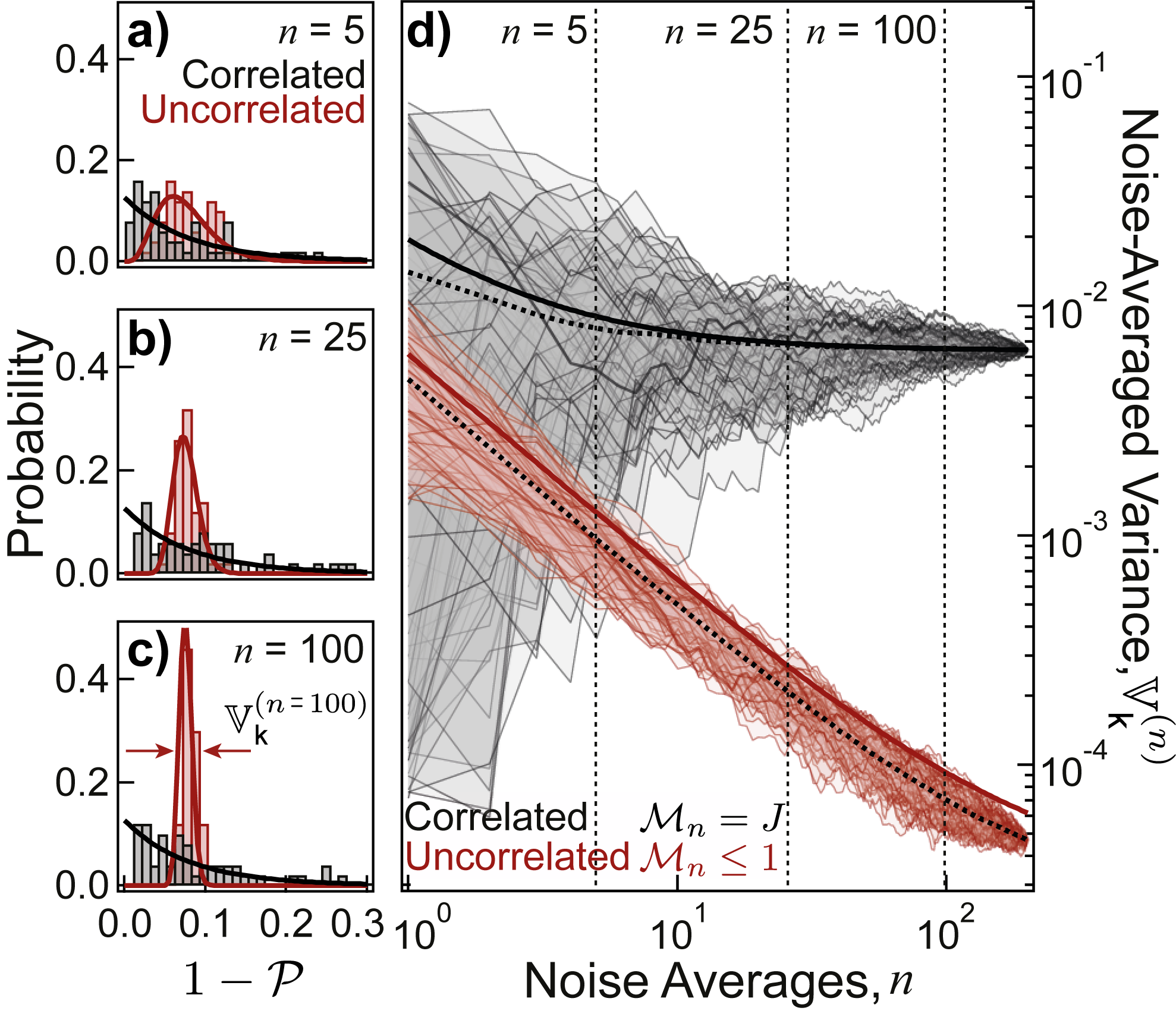}
	\caption{\textbf{Signatures of error correlations in randomized benchmarking sequences.}
\textbf{a-c,} Distribution of measured survival probabilities for $k=50$ randomly composed sequences averaged over $n=5, 25$ and 100 noise realizations drawn from \mbox{$\delta\sim\mathcal{N}(0,\rho^2=2\times10^{-3})$} for both maximally correlated, $\Mn = J$, (gray) and uncorrelated (red) engineered noise processes. Uncorrelated noise possesses a ``$\pi/2$-bandwidth'', i.e., noise values change with a rate set at the inverse of the duration of a primitive $\pi/2$-rotation, and hence can take one or multiple values in a gate ($\Mn \leq 1$). Solid lines are normalized Gamma distributions plotted with no free parameters. \textbf{d,} Scaling of cumulatively noise-averaged histogram variances, \mbox{$\mathbb{V}_\mathrm{k}^{(n)}  \equiv  \mathbb{V}_\mathrm{k}\left[ \noiseAve{P(\ket{1})} \right]$}. Trajectories correspond to different orderings of noise realizations with dotted lines representing the mean of 1000 re-orderings, and solid lines are theoretical predictions with no free parameters (see main text). Vertical dashed lines indicate the values of $n$ used in panels \textbf{a-c}.  
}
	\label{fig:engineerednoise}
\end{figure}

Figs.~\ref{fig:engineerednoise}a-c show the distributions over randomized benchmarking sequences of measured noise-averaged survival probabilities in the presence of concurrent detuning noise.  The same set of sequences is subject to correlated (gray) or uncorrelated (red) noise sampled from a common distribution. Data are represented as histograms for different fixed values of averaging number, $n$, for each sequence.  Solid lines are theoretical predictions for the distribution of survival probabilities derived from the updated random-walk framework, as given by the Gamma distributions from Eq.~\eqref{eq:gamma_dists}, and substituting the error rms value $\sigma$ using the noise-to-error translation for the expectation value shown in Table~\ref{tab:noise_to_error}. These theoretical predictions - which involve no free parameters -  show good agreement with the data in the regimes studied.  

These data clearly illustrate the differences in the distributions over the same set of randomized benchmarking sequences when subjected to noise with differing correlation properties. As shown in Ref.~\cite{Ball:2016} and highlighted here in Table~\ref{tab:statistical_moments}, the distributions possess approximately the same mean value, despite the differing noise-correlation properties.  The skew to high fidelities in the data taken using correlated noise is a manifestation of the randomized decoupling effects known to exist within some randomized benchmarking sequences~\cite{Ball:2016}. More importantly, the behavior of the variance of the distributions under an increasing number of noise averages $n$ varies substantially.  For small $n$ the distributions are similarly broad despite the differences in their shapes, but with further averaging the distribution measured under uncorrelated noise narrows while the variance of the distribution measured under correlated noise remains approximately constant (as discussed in Sec.~\ref{sec:2D_walks}). 

To highlight the effect of noise correlations on the experimental averaging behavior, we plot the variance of the distribution over measured sequence survival probabilities, \mbox{$\mathbb{V}_\mathrm{k}^{(n)}  \equiv  \mathbb{V}_\mathrm{k}\left[ \noiseAve{P(\ket{1})} \right]$}, as a function of the number of noise averages $n$ (Fig.~\ref{fig:engineerednoise}d).  Potential unintended systematic bias in the scaling of the experimental data with $n$ is mitigated by random re-ordering of the measured outcomes prior to cumulative averaging, producing a collection of individual averaging trajectories. For correlated noise, $\Mn=J$, the resulting trajectories are initially broadly distributed and fluctuate before converging with $n$ to a fixed, analytically calculable variance. By contrast, in the case of uncorrelated noise with $\Mn\leq 1$, all trajectories show an approximate reduction in \mbox{$\mathbb{V}_\mathrm{k}^{(n)} \propto 1/n$}, commensurate with a continued narrowing of the distribution of outcomes over different sequences under averaging \mbox{(Fig.~\ref{fig:engineerednoise}a-c)}. 

Solid lines capturing key scaling behaviors observed in both data sets of Fig.~\ref{fig:engineerednoise}d are derived from the expression for variance in Table~\ref{tab:statistical_moments} using the noise-to-error translations presented in Tables~\ref{tab:noise_to_error} and \ref{tab:expected_error_steps}, calculated for concurrent detuning noise with no free parameters. Overall, agreement with the measured experimental data are good across a wide parameter range and two orders of magnitude in $\mathbb{V}_\mathrm{k}^{(n)}$.  For correlated noise, small deviations between the theoretical trace and measured mean scaling appear for low values of $n$. Numerical evidence attributes this to the limited sample size in terms of sequences, which does not always capture the rare, highly error-susceptible sequences that would lead to a larger variance. In the case of uncorrelated noise, there is an overall vertical shift between the theory and the data, which is fully compensated by adjusting the rms noise strength $\rho_U$ by $\sim6\%$. Numerical simulations and analytic considerations attribute the need for this adjustment to the strong noise employed in these experiments, which violates the theoretical assumption $J\rho_U^2 \ll 1$, such that higher-order terms in the theory cannot be fully ignored.

The uncorrelated noise data begin to deviate from an exact $1/n$-scaling of $\mathbb{V}_\mathrm{k}^{(n)}$ at large numbers of noise averages. This behavior is captured by our theoretical model and varies in a predictable way with the applied noise bandwidth and sequence length $J$ (Appendix~\ref{app:expr_calculation}); we have verified it is not due to fundamental measurement limits in our system or quantum projection noise, as discussed in Appendix~\ref{app:QPN}. We are able to attribute this ``saturation'' in variance scaling for uncorrelated noise to residual sequence dependence, even in the case of purely uncorrelated noise, and the fact that our projective measurement probes only a two-dimensional $\sigx\sigy$-plane in Pauli-error space. For example, one can imagine a sequence composed solely of $\Id$ gates, which, due to an induced off-resonance error, will experience a net phase rotation that cannot be measured by single-axis projective measurements. Hence, no amount of averaging over different noise strength realizations will produce a survival probability that converges to the distribution mean, even in the case of uncorrelated noise.

Overall we find that our theoretical models predict not only the full distribution of survival probabilities over randomized benchmarking sequences, but also the scaling of this distribution's variance with experimental averaging.  The difference between the gray and red data in Fig.~\ref{fig:engineerednoise}d, and the agreement of theory, thus constitute key experimental validations of the central theoretical contributions made in this manuscript.

\section{Suppressing Error Correlations Using Dynamically Corrected Gates}
In the next part of our study we explore the ability to modify error correlations within a sequence through deterministic replacement of each Clifford operation in a randomized benchmarking sequence with an error-suppressing dynamically corrected gate (DCG).  Each DCG is implemented by replacing primitive physical rotations with composite pulses comprising multiple physical rotations~\cite{Kabytayev2014}, according to one of several prescriptions~\cite{True}.  This approach abstracts the target state transformations away from the physical qubit manipulation in a manner that builds in error robustness via coherent averaging. In this way, these composite gates modify the error susceptibility of the target operations, and in particular change the relationship between an input correlated-noise process and output gate errors. We therefore refer to their action as ``virtualizing'' the Clifford operations, consistent with an abstraction above the physical-layer operations presented in \cite{JonesPRX2012}.

\begin{figure*}[t!]
	\centering
	\includegraphics[scale=1]{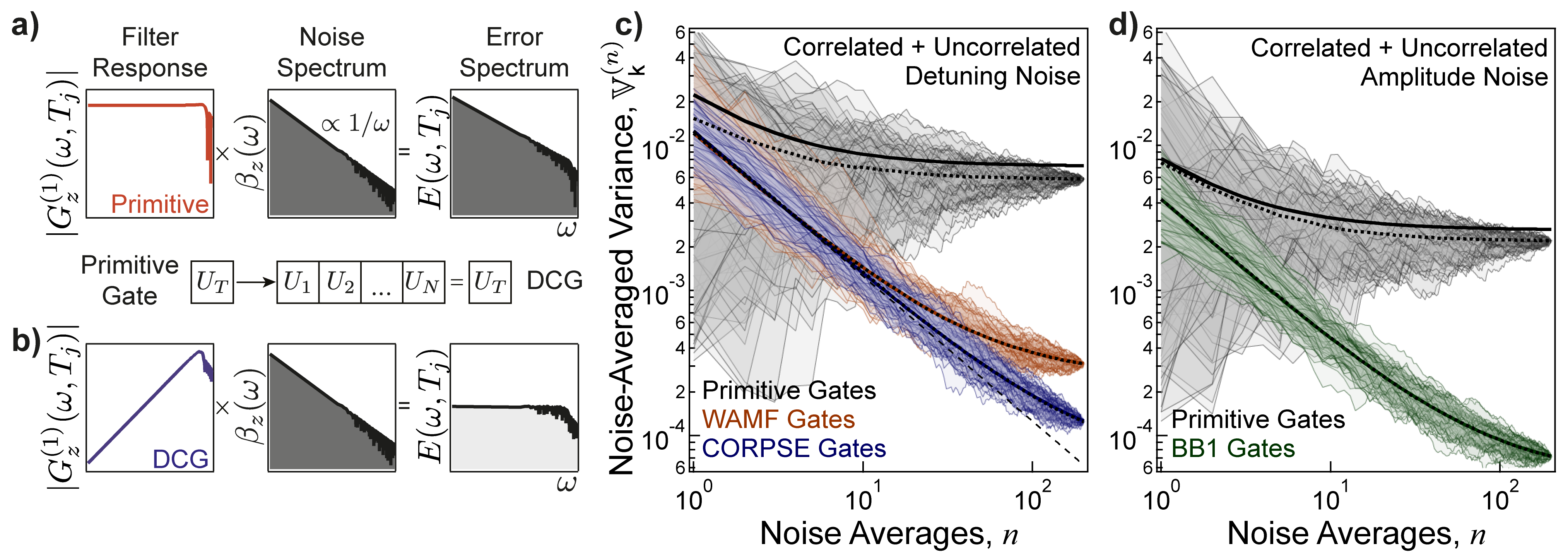}
	\centering
	\caption{
	\textbf{Suppression of error correlations using dynamically corrected gates.} 
	\textbf{a,} The first-order, generalized filter-transfer function for dephasing noise of a primitive operation $G^{(1)}_{z}(\omega, T_{j})$ and the noise spectrum (here $\beta_{z}(\omega)\propto1/\omega$) combine to produce an effective error spectrum $E(\omega, T_{j})$ for a single gate.   
    \textbf{b,} The modified filter functions for first-order DCGs scale as $\omega$ at low frequencies, which results in a ``whitening'' of $E(\omega, T_{j})$ relative to the input noise spectrum. 
    \textbf{c, d,} Variance scaling with $n$ for primitive (gray) gates, and WAMF (orange), CORPSE (blue), and BB1 (green) DCGs all subjected to noise with both correlated and uncorrelated components. For \textbf{c}, detuning noise is engineered with strength \mbox{$\delta_{C} \sim \mathcal{N}(0, 2\times10^{-3})$},  \mbox{$\delta_{U} \sim \mathcal{N}(0, 5\times10^{-4})$}, and for
    \textbf{d}, amplitude noise is engineered with strength \mbox{$\delta_{C} \sim \mathcal{N}(0, 9\times10^{-4})$},  \mbox{$\delta_{U} \sim \mathcal{N}(0, 2\times10^{-4})$}. Dotted lines are means of 1000 trajectories randomized over noise realizations, and solid lines for the DCGs are theoretical fits from Table~\ref{tab:statistical_moments} to the mean with the values of $\sigUncorr^2$ and $\sigCorr^2$ allowed to vary. Black solid lines for primitive gates are derived from the same theory with no free parameters. As with Fig.~\ref{fig:engineerednoise}, all data is measured for $k=50$ sequences of length $J=100$ with $n=200$ noise realizations and $r=220$ repetitions.
}
	\label{fig:DCG_engineerednoise}
\end{figure*}

The error-virtualization process is described quantitatively by calculating the error vector $\errorVect_j$ at the operator level and expressing it in the Fourier domain.  In the limit of classical Gaussian dephasing noise, described in the Fourier domain as the spectrum $\beta_{z}(\omega)$, the leading-order Magnus term ($\alpha=1$) in the $\sigz$-quadrature may be written as
\begin{equation}
    [\epsilon_{j, z}]_{1}=-i\int \frac{d\omega}{2\pi}G^{(1)}_{z}(\omega, T_j)\beta_{z}(\omega).
\end{equation}
Here, $G^{(1)}_{z}(\omega, T_j)$ is an analytically calculable, filter-transfer function that describes the spectral characteristics of a gate active for duration $T_j$~\cite{ViolaFFF}.  The \textit{effective} error spectrum experienced by the gate may therefore be represented by the spectral overlap of the filter-transfer function with the noise, written as \mbox{$G^{(1)}_{z}(\omega, T_j)\times \beta_{z}(\omega)\to E(\omega, T_j)$}. Fig.~\ref{fig:DCG_engineerednoise}a demonstrates the mapping between input noise and the effective error spectrum schematically for an example $1/\omega$-noise spectrum and a primitive $\pi$-rotation about the $x$-axis.  In this example, correlations in the noise are directly transferred to the correlations in the effective error spectrum~\cite{GreenPRL2012} (c.f. direct $\Mn$ to $\Me$ translation for primitive gates in Fig.~\ref{fig:noise2errorsRW}c).

Replacement of the primitive gate with a logically equivalent DCG virtualizes the effective error spectrum for each operator through the process of noise filtering~\cite{GreenPRL2012, Kabytayev2014, SoareNatPhys2014, ViolaFFF}.  Fig.~\ref{fig:DCG_engineerednoise}b illustrates this effect, where the DCG's reduced susceptibility to low frequency noise (captured through its filter-transfer function) results in a whitening of the effective error spectrum relative to $\beta_{z}(\omega)$.  In the current context, this whitening suggests that DCGs should not only reduce overall error magnitudes when the noise is dominated by low frequency contributions, but they should also suppress the signatures of error correlations between sequentially applied gates.

The particular DCG constructions examined in this work are the \enquote{Compensation for Off-Resonance with a Pulse SEquence} (CORPSE) \cite{Cummins:2000} and \enquote{Walsh Amplitude Modulated Filter} (WAMF) \cite{Ball:2014} gates, which suppress detuning errors, and the BB1 pulse family \cite{Wimperis:1994}, which suppresses over-rotation errors.  Specific details of DCG construction for the various operations employed here are presented in Appendix~\ref{app:DCG}.

\subsection{Modification of variance scaling with engineered errors using DCGs}

We begin by performing a detailed, quantitative study of the measured signatures of error correlations through the application of engineered noise. We experimentally implement primitive, CORPSE, WAMF and BB1 gates, where the first two DCGs are designed to suppress errors arising from frequency detuning noise and the latter is designed to suppress errors arising from amplitude noise. Using the same set of randomly generated randomized benchmarking sequences as in Fig.~\ref{fig:engineerednoise}, we now apply a mixed noise spectrum, simultaneously containing uncorrelated, rapidly varying noise \mbox{($\Mn \leq 1$)}, and quasi-static offsets that are constant over a full sequence giving a strongly correlated component \mbox{($\Mn = J$)}. In addition to performing measurements with primitive gates, we also construct DCG sequences by deterministically replacing each Clifford with its logically equivalent DCG counterpart. The relations for the mixed noise spectrum provided in Tables~\ref{tab:statistical_moments} and \ref{tab:noise_to_error} now permit a direct study of the impact of using DCGs on error correlations appearing within the randomized benchmarking sequences via the averaging behavior of $\mathbb{V}_\mathrm{k}^{(n)}$.

Beginning with frequency detuning noise, both DCG implementations shown in Fig.~\ref{fig:DCG_engineerednoise}c exhibit an initial variance scaling with noise averaging \mbox{$\mathbb{V}_\mathrm{k}^{(n)}\propto 1/n$}, reminiscent of the application of the purely uncorrelated noise process in Fig.~\ref{fig:engineerednoise}d.  The observed saturation in \mbox{$\mathbb{V}_\mathrm{k}^{(n)}$} at large $n$ for the DCG data combines contributions due to both the analytically calculable component occurring in the presence of purely uncorrelated noise introduced above, and residual uncompensated error correlations. The general behavior observed for the DCG sequences is to be contrasted with that observed for the same sequences composed of primitive gates where, as in Fig.~\ref{fig:engineerednoise}, the strong correlated noise component causes the variance to converge to a large constant value (gray).

Similar behavior is observed when considering the amplitude error quadrature. We demonstrate this through the application of engineered control-amplitude noise in Fig.~\ref{fig:DCG_engineerednoise}d, where measurements on sequences composed of DCGs derived from the BB1 family exhibit a similar \mbox{$\mathbb{V}_\mathrm{k}^{(n)}\propto 1/n$} averaging behavior. Again, this is contrasted with the behavior of sequences composed of primitive gates where once more the variance saturates to a high constant value, despite application of the same noise in both settings.

\subsection{Quantitative analysis of error-correlation suppression}

In order to calculate the change in error correlations realized in randomized benchmarking sequences composed of DCGs, we compare experimental measurements of \mbox{$\mathbb{V}_\mathrm{k}^{(n)}$} with the predictions of the model summarized in Table~\ref{tab:statistical_moments}. For the primitive gates, we explicitly translate the applied detuning noise strengths to an effective error strength using the noise-to-error relations in Table~\ref{tab:noise_to_error}; for this, we also use the expected random walk step expressions calculated and presented in Table~\ref{tab:expected_error_steps} of Appendix~\ref{app:revised_RB_theory} for detuning or amplitude noise with a $\pi/2$-bandwidth in the uncorrelated component. The solid, black lines in \mbox{Figs.~\ref{fig:DCG_engineerednoise}c,d} are then derived using these calculated error strengths, with no free parameters. Agreement between experimental measurements and theoretical predictions for the primitive gate sequences is good, but we observe a small ($\sim$20\%) deviation that appears approximately constant over several orders of magnitude in $n$ for both noise processes. Ongoing work is investigating the source of this discrepancy; possible sources include the unaccounted impact of higher-order terms due to the strength of the applied noise, and undersampling of the distribution over noise-averaged sequences.

To extract the relative correlated and uncorrelated error components after DCG application, we fit the data using the theoretical predictions for the scaling of $\mathbb{V}_\mathrm{k}^{(n)}$ shown in Table~\ref{tab:statistical_moments}, and use the strengths of the two error components $\sigUncorr^2$ and $\sigCorr^2$ as free parameters. First, for all DCGs we observe a reduction in $\sigCorr^2$ coupled with an increase in $\sigUncorr^2$.   Specifically, $\sigCorr^2$ is reduced by a factor of $49\times$ for CORPSE, $6\times$ for WAMF, and $10\times$ for BB1, while all experience an increase in $\sigUncorr^2$ by approximately $6-7\times$.  The relative performance of the DCGs observed in our experiments is aligned with their documented strengths, as CORPSE is known to more efficiently cancel purely static detuning errors than WAMF~\cite{Kabytayev2014, Ball:2014}, although improved calibration of the pulse-amplitude values used in WAMF gates is expected to improve the efficacy of correlated-error suppression.

The increase in $\sigUncorr^2$ is approximately consistent with the increase in duration of the DCGs relative to the primitive gate implementations. Considering the high-pass-filtering nature of all DCGs illustrates why uncorrelated noise processes fluctuating rapidly on the scale of the individual DCGs are transmitted by their filters and lead to residual errors that may be amplified by the DCG structure.  Overall, these measurements -- in particular the scaling of $\mathbb{V}_\mathrm{k}^{(n)}$ -- are consistent with an interpretation that the action of the noise whitening in the filter-transfer-function framework transforms correlated noise into predominantly uncorrelated residual errors \emph{at the operator level}.

\subsection{Signatures of variable error-correlation lengths}

\begin{figure}[t!]
	\centering
	\includegraphics[scale=1]{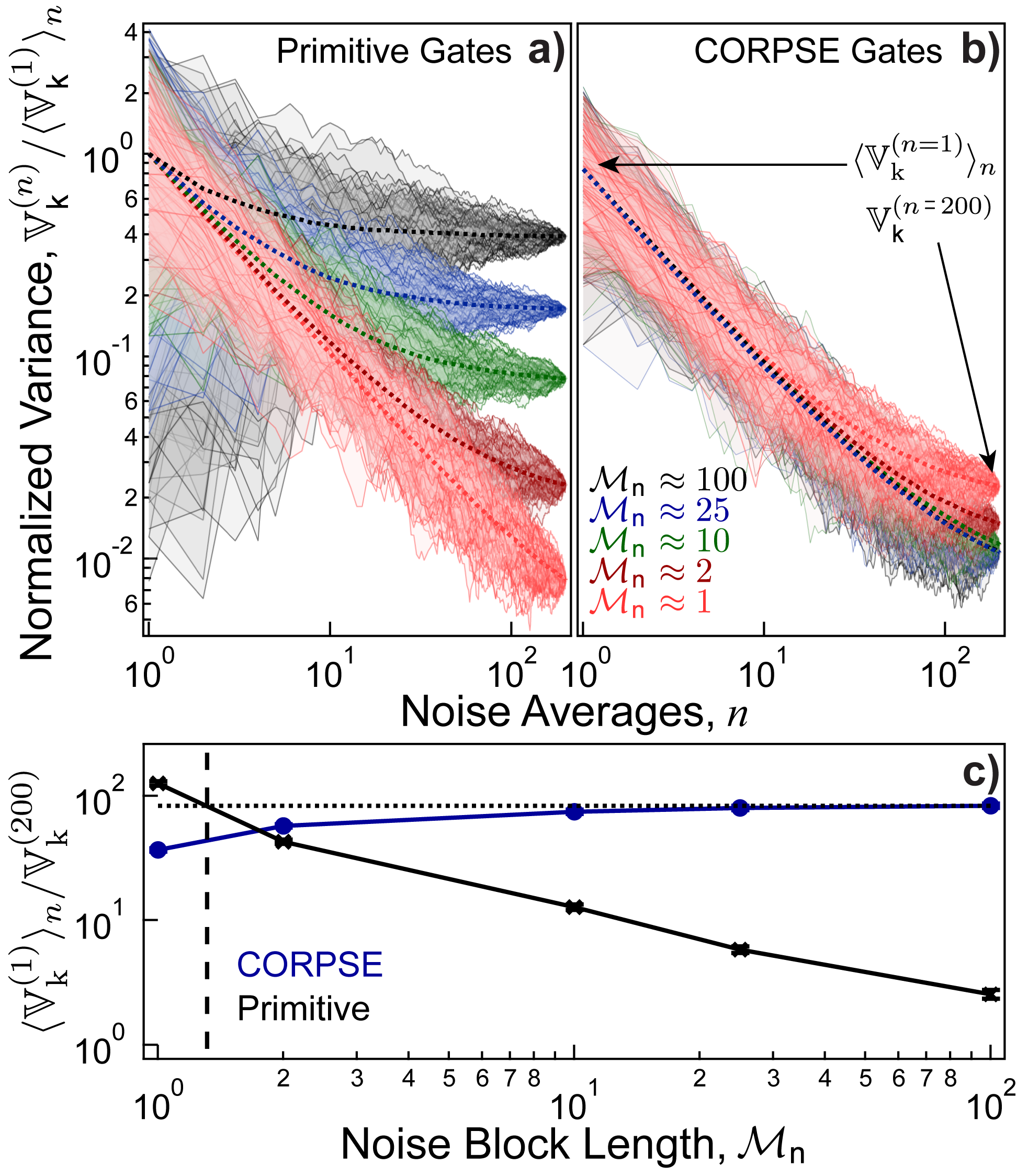}
	\caption{\textbf{Suppression of error correlations using DCGs under noise with varying $\Mn$.} \textbf{a, b,} Variance scaling of $k=20$ sequences with noise averaging for \textbf{a,} primitive and \textbf{b,} CORPSE gates. Traces are normalized to the initial mean variance for each applied noise case. Engineered noise is composed of an uncorrelated component ($\Mn \leq 1$) and a block correlated component of length $\Mn$ that is varied from fully correlated ($\Mn = J$) to uncorrelated ($\Mn = 1$) in units of virtual gates. Dotted lines are means of 1000 randomized trajectories. \textbf{c,} Ratio of initial to final variance in the upper panels as $\Mn$ is varied for primitive (black) and CORPSE (blue) gates. Dotted line marks the ratio at which CORPSE gates saturate, and the dashed vertical line indicates the value of $\Mn$ where this ratio crosses the scaling trend for primitive gates. Error bars calculated from the SEM of the 200 initial values of variance and normalized by the fully noise-averaged variance are smaller than point size.} 
	\label{fig:block_correlated_noise}
\end{figure}

To expand on the previous analyses, we experimentally demonstrate that the reduction in effective error correlation, indeed, resides at the virtual gate layer. Using the same sequences as before, and the same engineered $\rhoUncorr$ and $\rhoCorr$ rms magnitudes for detuning noise, the length of the correlated noise component is now varied in terms of the number of gates at the virtual level, breaking it up into blocks of length $\Mn$. The lab-frame durations of the noise blocks therefore now differ by a factor of $\sim6$ between the primitive and the CORPSE gates (the average increase in the duration of the Clifford operations when using CORPSE).

In the case of sequences composed of primitive gates, the signature exhibited by the variance scaling under noise averaging in Fig.~\ref{fig:block_correlated_noise}a gradually changes from indicating correlated errors (saturation at high variance) to purely uncorrelated errors ($1/n$-like scaling) as the block length is decreased, consistent with observations in Fig.~\ref{fig:engineerednoise} and Fig.~\ref{fig:DCG_engineerednoise}. By contrast, the sequences composed of CORPSE gates in Fig.~\ref{fig:block_correlated_noise}b retain their overall $1/n$-like scaling behavior for all correlated component block lengths, demonstrating that residual uncorrelated errors remain dominant. All traces in \mbox{Fig.~\ref{fig:block_correlated_noise}a,b} have been normalized to the initial mean variance for each engineered noise case to highlight the change in the relative correlated and uncorrelated error components, rather than the net error strength.

As a witness of the suppression of error correlations, Fig.~\ref{fig:block_correlated_noise}c shows the ratio of the initial mean variance $\mathbb{V}_\mathrm{k}^{(n=1)}$ to the final, fully noise-averaged variance $\mathbb{V}_\mathrm{k}^{(n=200)}$. This ratio scales approximately inversely with $\Mn$ for primitive gates but remains nearly constant for CORPSE gates. Extrapolation of this ratio for CORPSE back towards small $\Mn$ reveals a crossover with the primitive data that lies between $\Mn \approx 1$ to 2. This shows that CORPSE gates can reduce the noise correlation length to an error correlation length commensurate with physical noise $\Mn \approx 1$ to 2.  Because the noise correlation blocks were matched to the duration of the underlying Cliffords - whether through primitive or composite construction - these data highlight the efficacy of DCGs in virtualizing error characteristics for the logical gates implemented. 

\section{DCG's impact on intrinsic errors}

After verifying the utility of the theoretical constructs we have introduced in this work, we now turn to characterizing the intrinsic errors limiting the performance of our system. In the trapped $\Yb$ ion experiment described in Section~\ref{sec:experimental_setup}, we achieve a single-qubit randomized benchmarking average error per gate (EPG) of $(1.89 \pm 0.12) \times 10^{-5}$ (Appendix~\ref{app:single_qubit_RB}). Increasing the number of qubits to five and performing simultaneous randomized benchmarking using a global microwave control field reveals a monotonic increase in the EPG across the register, ranging from $(5.7 \pm 0.5) \times 10^{-5}$ to $(1.3 \pm 0.1) \times 10^{-4}$. As such, were we to run multi-ion algorithms that use global state manipulations, e.g., transversal gates in the 7-qubit Steane code~\cite{Steane:1996b}, we would not see the net error rate scale linearly with respect to the initial single-qubit EPG. This non-linear scaling with increasing qubit numbers has been observed in many systems and is often due to cross-talk between qubits~\cite{Proctor:2019}. It is important to note that this experimental observation of inhomogeneous error rates also violates a common assumption on noise statistics made in studies of error correcting codes, namely that the noise is independent and \emph{identically-distributed} (iid).

In our case, the underlying cause of the observed error inhomogeneity is a sub-percent-level gradient in the amplitude of the microwave control field across the ion chain, caused by interference from metallic surfaces in the proximity of our in-vacuum antenna.  We also observe a small magnetic-field gradient across the qubit chain, such that both amplitude and detuning noise are present simultaneously. Spatially correlated errors have recently been studied in reference~\cite{Postler:2018}, wherein it is noted that previous studies of multi-qubit errors tend to assume either spatially independent errors or identically spatially correlated errors, facilitating the use of a decoherence free subspace. Our situation, with a gradient of spatially correlated errors, falls between these two cases, but can still induce simultaneous multi-qubit errors that lower the efficacy of QEC.

To characterize the impact of DCGs on spatially correlated errors, we utilize simultaneous randomized benchmarking sequences of length $J=500$ applied to all five qubits in the register, and again explore variance scaling with experimental averaging.  We construct DCG sequences using BB1 gates to combat the dominant microwave-control-amplitude errors. Data collection proceeds by interleaving a single sequence implemented using either primitive or BB1 gates to ensure a fair comparison between the sequences in time, in the event that any systematic drifts occur. 

\begin{figure}
	\centering
	\includegraphics[scale =1]{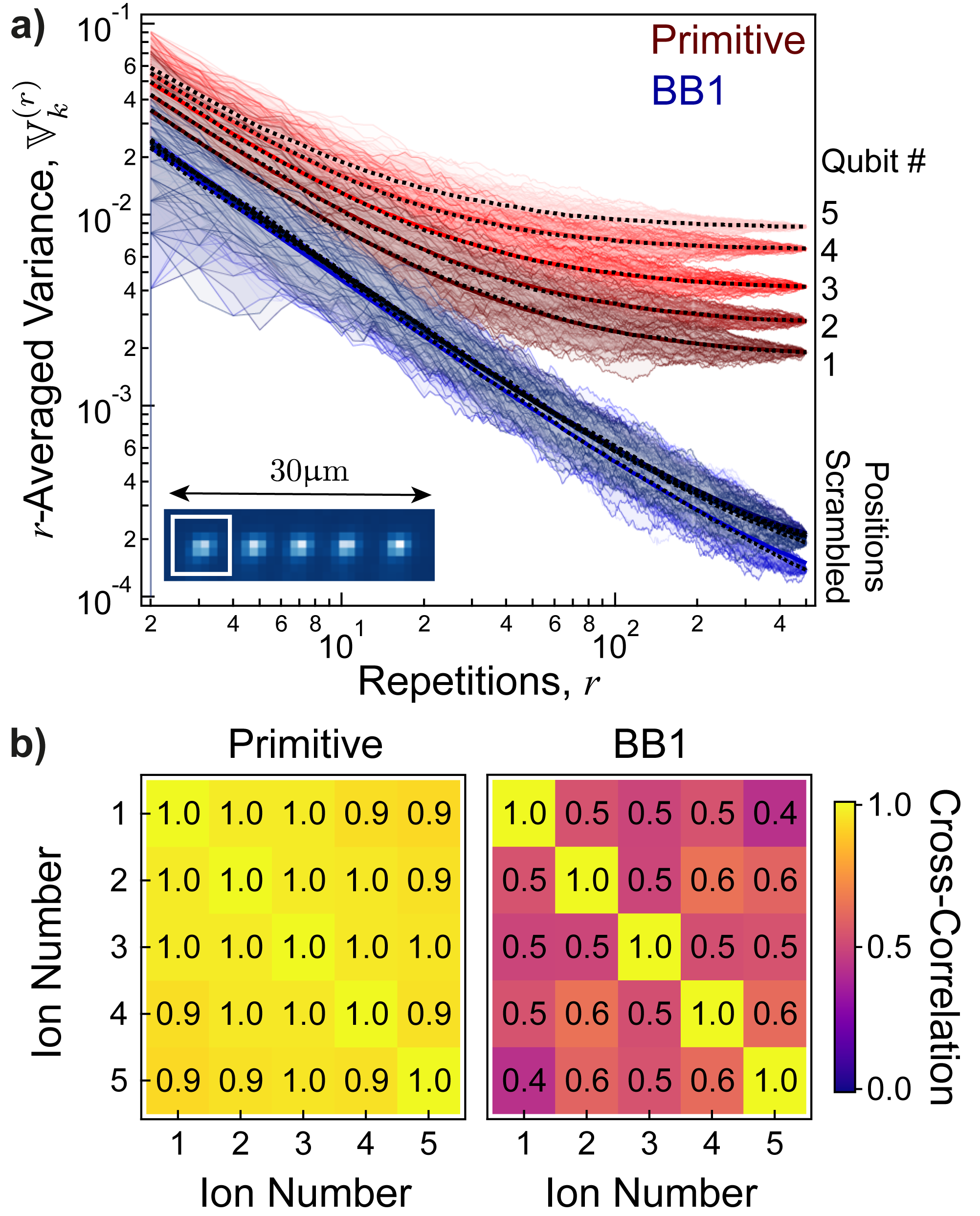}
	\caption{\textbf{Intrinsic errors in a five-qubit chain.} \textbf{a,} Variance over noise-averaged sequence survival probabilities for five-qubits using $k=60$ sequences of length $J=500$, averaged over repetitions $r$, up to $r=500$. Each trajectory is produced by shuffling the order of repetitions used in the graph to avoid bias, dotted lines indicate the means of 1000 trajectory randomizations, and solid lines are fits where the correlated and uncorrelated error strengths were free to vary. The correlated error strengths, $\sigCorr^2$, are $\{1.2, 1.5, 1.9, 2.4, 2.7\}\times10^{-4}$ from qubit 1 to 5 for the primitive gates, and $\{2.3, 2.5, 1.1, 2.2, 2.3\}\times10^{-5}$ for the BB1 gates. The uncorrelated error strengths, $\sigUncorr^2$, are $\{7.5, 8.1, 8.5, 8.6, 8.7\}\times10^{-4}$ from qubit 1 to 5 for the primitive gates, and $\{6.5, 6.5, 6.3, 6.6, 6.5\}\times10^{-4}$ for the BB1 gates. 
	(Inset) EMCCD image of a five-ion chain, spaced over $\sim 30 \upmu$m. The control field amplitude and frequency is calibrated with respect to the highlighted, leftmost ion.
\textbf{b,} Pairwise, cross-correlation coefficients between the five-qubit survival probabilities for primitive gates (left) and BB1 DCGs (right), revealing a $\sim50\%$ reduction in the correlations between qubit errors when using DCGs.}
	\label{fig:intrinsic_errors}
\end{figure}

We examine the scaling of $\mathbb{V}_{\mathrm{k}}^{(r)}$ with averaging over repetitions $r$, up to $r=500$; because noise is native to the system, we make the substitution $n\equiv r$. The signature of the temporally correlated intrinsic errors is observed for all ions when using sequences of primitive gates in Fig.~\ref{fig:intrinsic_errors}a (red).  We observe a staggered, increasing saturation value for $\mathbb{V}_{\mathrm{k}}^{(r)}$ at $r=500$, increasing with the spatial distance from qubit 1 (leftmost qubit in Fig.~\ref{fig:intrinsic_errors}a inset), which is used to calibrate the gate operations. As expected, the qubit that is furthest from the calibration qubit suffers both the worst randomized benchmarking performance and shows the highest saturation value in variance scaling. By contrast, the over-rotation error suppressing BB1 gates (blue) saturate at a value of variance over an order of magnitude lower than achieved by the primitive gates, and recover a $1/r$-like scaling for all qubits. We further find the relationship between the physical positions of the qubits and the ordering of saturation variances has become scrambled. Using the analysis introduced above, we fit the mean variance trends with the expression in Table~\ref{tab:statistical_moments}, allowing the strengths of the error $\sigCorr^2$, $\sigUncorr^2$ to vary. We extract a reduction in the correlated error strength when using BB1 gates ranging from $\sim5$ to $16\times$ for the five qubits.

To directly probe the action of DCGs in virtualizing the spatially correlated errors, we calculate the pairwise cross-correlation coefficient between the survival probabilities in each experimental realization (Fig.~\ref{fig:intrinsic_errors}b). For primitive gates, all errors are highly correlated between qubits (cross-correlation coefficient $\geq 0.9$ for all qubit pairs), whereas for the BB1 gates, a reduction of approximately 50\% can be seen between all qubit pairs, further supporting the evidence that DCGs provide a suppression of error correlations in both time and space.

Separate investigations not presented here using the multi-axis error suppressing DCG CinBB showed no additional benefit.  This observation suggests that the off-resonance error created by the magnetic-field gradient was sufficiently small that it was dominated by other larger, but rapidly fluctuating, intrinsic error sources

%
\section{Outlook}
%

The results we have presented suggest that the path to the practical implementation of QEC may be facilitated by transforming miscalibrations and common laboratory noise sources exhibiting slow drifts and low-weight noise spectra, into effective error processes with dramatically reduced correlations at the virtual layer using DCGs. We believe this is important as the pursuit of functional quantum computers -- even at the mesoscale -- will clearly require major advances in the control and suppression of errors, as gate counts quickly exceed $10^{10}$ for even moderate problems requiring only $\sim200$ qubits  \cite{Reiher:2017}. Combined with the observation that certain DCGs can mitigate spatial cross-talk in multi-qubit systems \cite{Merrill:2014}, we believe that our demonstration of the suppression of temporal and spatial error correlations within quantum circuits solidifies the central importance of dynamic error suppression techniques at the virtual level for practical quantum computing.

\section*{Acknowledgments}
The authors acknowledge S. Mavadia for assistance with data collection and simulations, and discussions with H. Ball, C. Ferrie, and C. Granade on data analysis. Work partially supported by the ARC Centre of Excellence for Engineered Quantum Systems CE170100009, the Intelligence Advanced Research Projects Activity (IARPA) through the US Army Research Office Grant No. W911NF-16-1-0070, and a private grant from H. \& A. Harley.

\appendix
\section*{Appendix}

\section{Single-qubit randomized benchmarking}
\label{app:single_qubit_RB}

\begin{figure}[ht]
	\centering
	\includegraphics[scale=1]{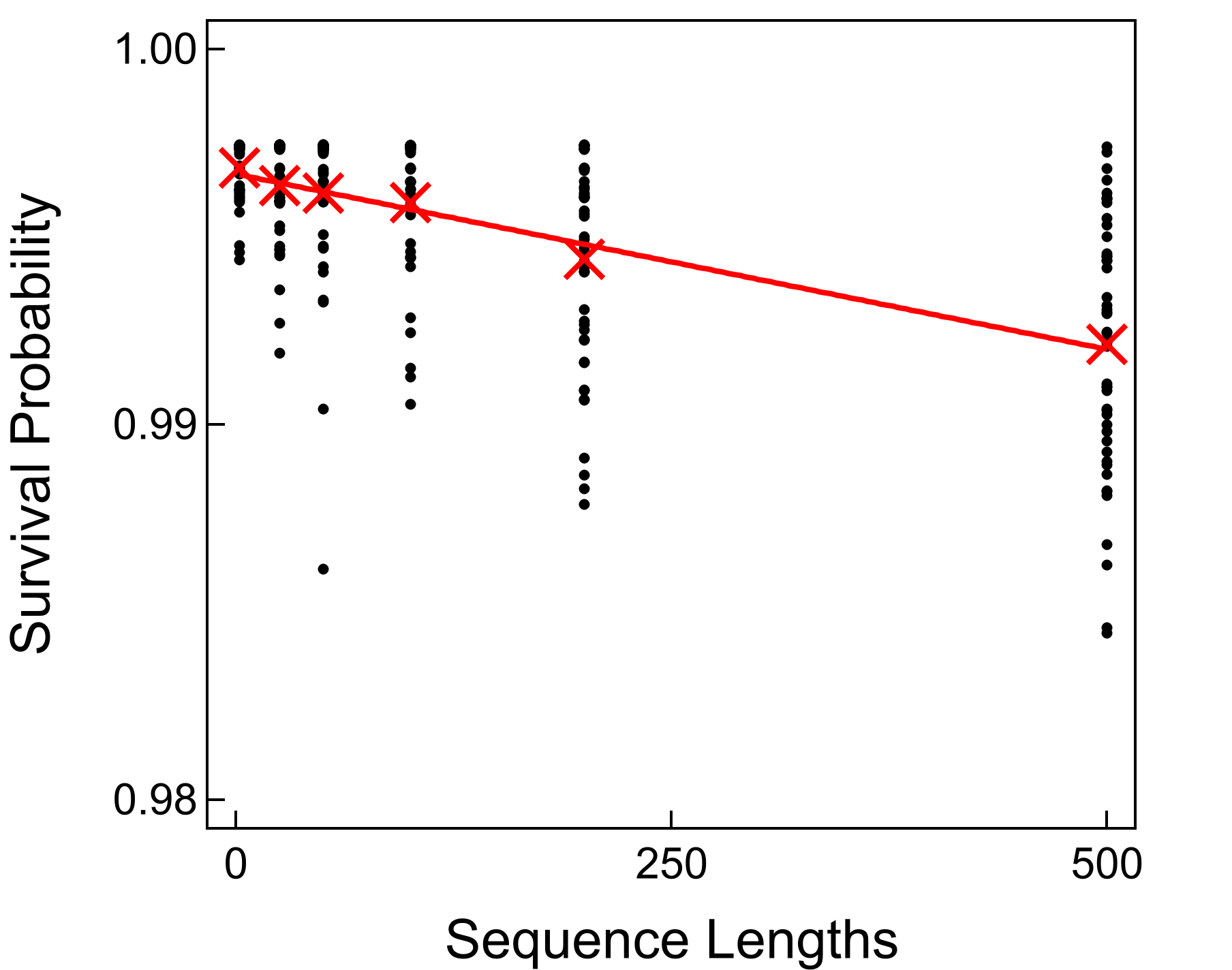}
	\caption{\textbf{Single-qubit randomized benchmarking.} Randomized benchmarking is performed on a single qubit using a total of 300 sequences composed of primitive gates, with 50 sequences each of length $J=2,\, 25,\, 50,\, 100,\, 200,$ and $500$. Each sequence was repeated $r=500$ times to reduce quantum projection noise. Black markers represent individual sequence survival probabilities, red crosses indicate the mean survival probabilities for each sequence length, and the solid red line in a fit to the means to extract the average error per gate.}
	\label{fig:single_qubit_RB}
\end{figure}

Using the experiment described in Section~\ref{sec:experimental_setup}, with a single trapped $\Yb$ ion and microwave gates, we achieve a single-qubit error per gate of \mbox{$p_\textrm{RB}=(1.89\pm0.12)\times 10^{-5}$} measured using randomized benchmarking (Fig.~\ref{fig:single_qubit_RB}). The fit to the mean survival probabilities used to extract the error per gate is given by
\begin{equation}
    \mathcal{P} = 0.5 + (0.5 - \kappa) \textrm{e}^{-p_\textrm{RB}J},
\end{equation}
where $\mathcal{P}$ is the mean survival probability, $J$ is the number of gates in a randomized benchmarking sequence, and $\kappa$ is the value of our single-qubit State Preparation and Measurement error (SPAM), found to be \mbox{$\kappa = (3.3 \pm 0.1) \times 10^{-3}$}.

\section{Physical noise to error strength translation}
\label{app:revised_RB_theory}

We verify the model presented in this manuscript by using primitive Clifford gates under engineered noise, where the strength and effect of the noise are known exactly, allowing for quantitative analysis. For this verification, we need to calculate the translation between the rms magnitude of the physical noise process, $\rho$, and that of the resulting error operators, $\sigma$. The noise is applied concurrent with the gate operations inducing multi-axis and gate-dependent errors for the different Clifford operations, whose lengths differ between $\pi$ and $\pi/2$ rotations. Due to this introduced gate-dependence, an exactly constant noise process will not be directly translated to a constant error process with identical error vectors for every gate, and hence the translation for each gate needs to be considered explicitly.

The method to transform noise strength to error strength for noisy, primitive Clifford gates is initially presented here for a general noise process that is static over the duration of a single gate. Each of the single-qubit Clifford gates are made up of rotations on the Bloch sphere with the rotation axis and angle specified by the Clifford gate index, $\etaVectj{j} \in \{1, \dots, 24\}$. If the $j$th gate in a sequence is affected by laboratory noise with value \mbox{$\delta_j\sim\mathcal{N}(0, \rho^2)$}, the resulting noisy gate can be decomposed into an error operator and the ideal gate, \mbox{$\noisyC{\etaVectj{j}}=\errorOp{j} \cleanC{\etaVectj{j}}$}, with
\begin{align}
\errorOp{j} 
&= e^{ i \sum_{\alpha=1}^\infty \delta_j^\alpha [\nuVect{\etaVectj{j}}]_\alpha  \cdot \sigVect }		 \\
&\approx \Id + i \delta_j [\nuVect{\etaVectj{j}}]_1 \cdot \sigVect	\nonumber
\end{align}
where $\sigVect$ is the vector of Pauli matrices. In the main text, this operator was introduced in terms of the error vector $\errorVect_j$ as \mbox{$\errorOp{j} = \exp \left\{ i \sum_{\alpha=1}^\infty [\errorVect_{j}]_\alpha  \cdot \sigVect \right\}$}. We have now separated the error vector into two components for the Magnus expansion of order $\alpha$, \mbox{$[\errorVect_j]_\alpha = \delta_j^\alpha [\nuVect{\etaVectj{j}}]_\alpha$}, to explicitly show the dependence on the physical noise strength $\delta_j$, which will change between different realizations of the noise, and the particular gate's susceptibility to the error channel, described by the term $\nuVect{\etaVectj{j}}$. There will be 24 gate-specific error vector terms, $\nuVect{\etaVectj{j}}$, corresponding to the 24 Clifford operations, which can be calculated explicitly for a given noise process. We now consider how these terms affect our ideal randomized benchmarking sequence.

Starting with the standard randomized benchmarking procedure, we compile a sequence of randomly composed single-qubit Clifford operations, \mbox{$\prod_{j=1}^J \cleanC{\etaVectj{j}} = \Id$}, which are mathematically right-multiplied to the preceding operator such that they act sequentially on an initial state. Then, the complete noisy sequence is given by
\begin{align}
\tilde{S} &= \prod_{j=1}^J \errorOp{j} \cleanC{\etaVectj{j}}.
\end{align}
The survival probability for a qubit prepared in $\ket{0}$, averaged over $n$ noise instances, is calculated using
\begin{equation}
\mathcal{P} = 1-\noiseAve{ P(\ket{1}) } = \noiseAve{ \lvert \bra{0} \tilde{S} \ket{0} \rvert^2}.
\end{equation}

To approximate the sequence, the method from \cite{Ball:2016} is employed: the first-order term of each error operator can be translated to a step in Pauli-error space, with the total random walk in three dimensions for a given noise instance $i$ given by
\begin{equation}
    \label{eq:twoD_walk}
    \Ri{\textrm{3D}}{i} = \stepSum{j} \del{j}{i} \rVect{\textrm{3D}, j}.
\end{equation}
The $j$th random walk step, $\rVect{\textrm{3D}, j}$, is calculated from the product of the preceding ideal gates modifying the first-order, gate-specific error for the $j$th operation in the sequence \mbox{$[\nuVect{\etaVectj{j}}]_1\cdot\sigVect$},
\begin{align}
\begin{aligned}
\label{eq:rvect_calc}
    \cleanC{\etaVectj{{1}}} \dots \cleanC{\etaVectj{{j-1}}} \left(  [\nuVect{\etaVectj{j}}]_1 \cdot \sigVect \right) (\cleanC{\etaVectj{{1}}} \dots \cleanC{\etaVectj{{j-1}}})^\dagger\\  = \rVect{\textrm{3D}, j} \cdot \sigVect.
\end{aligned}
\end{align}

To obtain the sequence survival probability that would be measured via a single-axis projective measurement, the relevant steps are then the projection of $\rVect{\textrm{3D}, j}$ in the two-dimensional $\sigx\sigy$-plane, \mbox{$\rVect{\textrm{2D}, j}\equiv\rVect{j}$}, of Pauli-error space. As with the original model, it can be shown that a sequence's survival probability is given by
\begin{align}
\mathcal{P} &= 1 - \noiseAve{\norm{\R{}}^2} +\Order{\rho^4},
\end{align}
where $\R{}$ is the two-dimensional random walk. From this expression, the expectation and variance of the distribution over noise-averaged sequence survival probabilities have been calculated for arbitrary step lengths; the results of this calculation are summarized in the noise-to-error translation in Table~\ref{tab:noise_to_error} of the main text. These expressions are based on the expected random walk steps induced by the 24 error maps, which are shown in Table~\ref{tab:expected_error_steps} for a range of physical noise processes. We proceed here by showing an example derived for a concurrent detuning error.

\begin{table*}[!t]
{\renewcommand{\arraystretch}{2}
\begin{tabular}{|c|c|c|c|}
\hline
	\textbf{Engineered Noise Process}
& 	$\exprTwo{j}$ 	  	
& 	$\exprFour{j}$ 	
& 	$\Cov{ \norm{\rVect{U, j}}^2 }{\norm{\rVect{C, j}}^2 }$
\\ \hline
	Interleaved Dephasing
& 	2/3
& 	2/3
&	2/9
\\ \hline
	\parbox[c][30pt][c]{3\textwidth/9}{Concurrent Detuning\\
	1 value per primitive ($\pi$ or $\pi/2$) gate}
& 	$\frac{2}{3} \left( \frac{1}{2} + \frac{\pi^2}{96} \right)$
& 	$\frac{2}{3} \left( \frac{7}{24} + \frac{\pi^4}{384} \right)$
&	$\frac{2}{3} \left( \frac{7}{24} + \frac{\pi^4}{384} \right)
      - \frac{4}{9} \left( \frac{1}{2} + \frac{\pi^2}{96} \right)^2 $
\\ \hline
	\parbox[c][40pt][c]{3\textwidth/9}{Concurrent Detuning\\
    1 value every primitive $\pi/2$ gate time}
& 	$ \frac{2}{3} \left( \frac{1}{2} + \frac{\pi^2}{192} \right)$
& 	$\frac{2}{3} \left( \frac{1}{4} + \frac{\pi^4}{1536} \right)$
&	$\frac{17}{108}+\frac{\pi ^4}{1152}
     - \frac{4}{9} \left( \frac{1}{2} + \frac{\pi^2}{192} \right) 
                    \left( \frac{1}{2} + \frac{\pi^2}{96} \right)$
\\ \hline
	\parbox[c][40pt][c]{3\textwidth/9}{Over- and Under-rotation\\
    1 value per primitive ($\pi$ or $\pi/2$) gate}
& 	$ \pi^2/18 $
& 	$ 5\pi^4/576 $
&	$ 29\pi^4/5184 $
\\ \hline
	\parbox[c][40pt][c]{3\textwidth/9}{Over- and Under-rotation\\
    1 value every primitive $\pi/2$ gate time}
& 	$ \pi^2/36 $
& 	$ 5\pi^4/2304 $
&	$ 29\pi^4/10368 $
\\ \hline
\end{tabular}}
\caption{The expected step lengths in the Pauli $\sigx\sigy$-plane based on the average Clifford gate error for different engineered noise. These quantities are used to predict the statistical moments of the sequence survival-probability distributions in Table~\ref{tab:noise_to_error}.}
\label{tab:expected_error_steps}
\end{table*}

\subsection{Example for Concurrent Detuning Noise} 
\label{app:expr_calculation}

The combined main text Tables~\ref{tab:statistical_moments} and \ref{tab:noise_to_error} predict the form of the noise-averaged survival-probability distribution for different engineered noise processes, given the expected random walk steps in the $\sigx\sigy$-place of Pauli-error space, $\exprTwo{j},\, \exprFour{j}$. As an explicit example, these quantities are calculated here for concurrently applied detuning noise, produced by an offset between the qubit frequency and the control field frequency, normalized to the Rabi frequency, \mbox{$\delta=\Delta/\Omega$}. An ideal rotation of angle $\theta$ about the $\n$-axis of the Bloch sphere is modified by detuning noise as,
\begin{equation}
\tilde{U}(\n, \theta, \delta)=e^{-i\left( \theta \sigVect\cdot\n +  \abs{\theta} \delta \sigz \right)/2}.
\end{equation}
From this, the eight physical error maps affecting the Clifford operations are calculated to be,
\begin{subequations}
\label{eq:decomposed_all}
\begin{align}
	\errorOpw{}{\Id}(\pi,\delta)&= \Id-i\tfrac{\pi\delta}{2}\sigz+ \Order{\delta^2}	\\
	\errorOpw{}{\sigx}(\pi,\delta)&= \Id  +i \delta \sigy+ \Order{\delta^2}	\\
	\errorOpw{}{\sigx}(\pm\tfrac{\pi}{2},\delta)&= \Id \pm\tfrac{i \delta }{2}\sigy-\tfrac{i \delta }{2}\sigz+ \Order{\delta^2} 	\\
	\errorOpw{}{\sigy}(\pi,\delta)&= \Id -i \delta \sigx + \Order{\delta^2}	\\
	\errorOpw{}{\sigy}(\pm\tfrac{\pi}{2},\delta)&= \Id \mp \tfrac{i \delta }{2}\sigx - \tfrac{i \delta }{2}\sigz+ \Order{\delta^2}	\\
	\errorOpw{}{\sigz}(\theta, \delta)&= \Id,
\end{align}
\end{subequations}
more generally expressed for the $j$th operation in the sequence as
\begin{equation}
	\errorOp{j} = \Id + \delta_j [\nuVect{\etaVectj{j}}]_1 \cdot \sigVect + \Order{\delta_j^2}.
\end{equation}
Only eight error maps are required to treat all 24 Clifford operations due to the error-free nature of $\sigz$-rotations, which are generally implemented via instantaneous phase-changes on the control field. Following the definition of the Clifford operations given in \cite{Ball:2016}, there is only one non-$\sigz$-rotation per Clifford, which exactly corresponds to one of the eight error maps described in \eqref{eq:decomposed_all}. If $\sigz$ operations were also affected by the noise, the procedure would follow similarly but all error maps would need to be calculated.

To find the expected random walk steps for this unitary error channel, recall from \eqref{eq:rvect_calc} that the direction of the Pauli-error steps is determined by the preceding operations in the randomly composed sequence. As such, a given step will remain deterministic in its size, yet be performed along an arbitrary direction in Pauli-error space, determined by the preceding gates. Studying the error maps for concurrent detuning noise, we can write the gate-dependent steps as 
\begin{align}
    \errorOpw{}{\Id}(\pi)  &\rightarrow \tfrac{\pi}{2}\hat{m}_1	\\
	\errorOpw{}{\sigx}(\pi)  &\rightarrow 1\hat{m}_1	\\
    \errorOpw{}{\sigx}(\tfrac{\pi}{2})  &\rightarrow(\tfrac{1}{2}\hat{m}_1 +  \tfrac{1}{2}\hat{m}_2)	\\	
	\errorOpw{}{\sigy}(\pi)  &\rightarrow 1\hat{m}_1	\\
	\errorOpw{}{\sigy}(\tfrac{\pi}{2}) &\rightarrow(\tfrac{1}{2}\hat{m}_1 +  \tfrac{1}{2}\hat{m}_2)	\\
	\errorOpw{}{\sigz}(\theta)  &\rightarrow 0
\end{align}
with $\hat{m}_1,\hat{m}_2 \in \pm\{\sigx,\sigy,\sigz\}$. This implies that $\pi$-rotations about the $x$ and $y$-axes of the Bloch sphere produce a unit-length step in Pauli-error space that will be randomly oriented along one of the six principal axes. Similarly, $\pi/2$-rotations produce a $1/\sqrt{2}$-length step oriented at $45^\circ$ between two principal axes, $\Id$ gates produce a $\pi/2$-length step along a principal axis, and rotations about the $z$-axis contribute no step due to their error-free nature. 

The probability of producing a particular non-zero $\modr{j}$ is shown in Table \ref{tab:modr_probs}, based on the prevalence of different gates in the 24 Clifford gates and the likeliness of their projection into the $\sigx\sigy$-plane. Note that these steps are completely independent of the strength of the particular noise realization, $\del{j}{i}$. The noise will eventually rescale each step length, but here we only consider the unscaled walk.
For this particular noise type and bandwidth, it is not necessary to distinguish between $\exprTwoCorr{j}$ and $\exprTwoUncorr{j}$, as both the correlated and uncorrelated error processes are static over the duration of an individual gate, and hence will result in the same expected average walk steps; it is only when increasing the bandwidth of the uncorrelated noise that they need be distinguished. Using Table \ref{tab:modr_probs} one finds
\begin{align}
\exprTwo{j} 
&= \frac{2}{3} \left( \frac{1}{2} + \frac{\pi^2}{96} \right) \\
\exprFour{j} 
&= \frac{2}{3} \left( \frac{7}{24} + \frac{\pi^4}{384} \right)\\
\exprTwoCorrUncorr{j} &= \exprFour{j}.
\end{align}
\begin{table}[t]
\center
{\renewcommand{\arraystretch}{1.5}
\begin{tabular}{|c|c|c|c|c|}
\hline
$\modr{j}$
& 	1
&	$\tfrac{1}{\sqrt{2}}$
&	$\tfrac{1}{2}$
&	$\tfrac{\pi}{2}$
\\	\hline
$\textrm{Pr}_{\sigx\sigy}$	
&  $\tfrac{4}{24}\times\tfrac{2}{3}$
& 	 $\tfrac{16}{24}\times\tfrac{4}{12}$	
& 	 $\tfrac{16}{24}\times\tfrac{8}{12}$	
&	$\tfrac{1}{24}\times\tfrac{2}{3}$	
\\ \hline
\end{tabular}}
\caption{Likelihood of producing particular length random walk steps in the $\sigx\sigy$-plane of Pauli-error space when engineered detuning noise is applied, based on the number of Clifford gates corresponding to the error map, and the chance of a randomly oriented step in the $\sigx\sigy$-plane.}
\label{tab:modr_probs}
\end{table}
Using Tables~\ref{tab:statistical_moments} and \ref{tab:noise_to_error}, this produces the expectation value
\begin{equation}
	\E{\mathcal{P}}  \approx J \sigma^2 \tfrac{2}{3}\left(\tfrac{1}{2}+\tfrac{\pi^2}{96}\right) \label{eq:meanstdc}
	\end{equation}
for both correlated and uncorrelated errors. This again illustrates the equivalence of the distribution mean, which is related to the parameter that standard randomized benchmarking analysis returns, for noise of the same strength despite vastly different correlation lengths. The difference between the correlated and uncorrelated processes becomes evident when looking at the variance over survival probabilities with increased noise averaging.

For uncorrelated errors,
\begin{align}
\Var{\mathcal{P}_U} 	\approx&	\tfrac{J^2 \sigma^4}{n}\left[ \tfrac{4}{9}\left(\tfrac{1}{2}+\tfrac{\pi^2}{96}\right)^2+\tfrac{1}{J}\left\{ 3\left(\tfrac{7}{36}+\tfrac{\pi^4}{576}\right)\nonumber \right.\right.\\
&\left.\left.-\tfrac{8}{9}\left(\tfrac{1}{2}+\tfrac{\pi^2}{96}\right)^2\right\} +\tfrac{(n-1)}{J}\left\{ \tfrac{7}{36}+\tfrac{\pi^4}{576}\right.\right.\nonumber \\
&\left.\left.-\tfrac{4}{9}\left(\tfrac{1}{2}+\tfrac{\pi^2}{96}\right)^2\right\}  \right], \label{eq:varstac}
\end{align}
noting that in the limit $n\rightarrow\infty$, the variance scaling saturates at a value $\propto \tfrac{1}{J}$ relative to the starting variance. For correlated errors,
\begin{align}
\Var{\mathcal{P}_C}	\approx&	\tfrac{J^2 \sigma^4}{n}\left[ \tfrac{12}{9}\left(\tfrac{1}{2}+\tfrac{\pi^2}{96}\right)^2+\tfrac{1}{J} \left\{ 3\left(\tfrac{7}{36}+\tfrac{\pi^4}{576}\right)\nonumber \right.\right.\\
&\left.\left.-\tfrac{8}{3}\left(\tfrac{1}{2}+\tfrac{\pi^2}{96}\right)^2\right\} +(n-1)\left\{ \tfrac{4}{9}\left(\tfrac{1}{2}+\tfrac{\pi^2}{96}\right)^2+\nonumber \right.\right.\\
&\left.\left.\tfrac{1}{J}\left(\tfrac{7}{36}+\tfrac{\pi^4}{576}-\tfrac{8}{9}\left(\tfrac{1}{2}+\tfrac{\pi^2}{96}\right)^2\right)\right\} \right], \label{eq:varstdc}
\end{align}
again tending towards a constant; however, this occurs at a significantly smaller number of noise averages than for uncorrelated noise and saturates at a much larger variance $\propto 1 + \tfrac{1}{J}$ relative to the starting variance.

Using the revised model, the noise-averaged survival-probability distributions under correlated noise remain Gamma distributed with an updated scale parameter. 
While this is yet to be shown explicitly for the uncorrelated case, the behavior is approximated in the limit of large $n$ and $J$, with $n<J$ by modifying the distribution in \eqref{eq:gamma_dists}, to yield 
\begin{align}
\mathcal{P}_C &\sim \Gamma(a = 1, b = \tfrac{2}{3}J\sigma^2 (\tfrac{1}{2}+\tfrac{\pi^2}{96}) ),  \label{eq:2DCorOnConGamma}  \\
\mathcal{P}_U &\sim \Gamma(a = n, b = \tfrac{2}{3n}J\sigma^2  (\tfrac{1}{2}+\tfrac{\pi^2}{96}) ) \label{eq:2DUncorOneConGamma}.
\end{align}
The normalized Gamma distributions for correlated error processes shown by solid gray lines in the main text Fig~\ref{fig:engineerednoise}a-c were calculated from first principles using \eqref{eq:2DCorOnConGamma} with no free parameters. The distributions for the uncorrelated error process in red were calculated from an altered version of \eqref{eq:2DUncorOneConGamma}, which was modified for higher bandwidth noise that took multiple values of $\delta$ in a single error map. This made use of the relation
\begin{align}
\delta_1 \pm  \delta_2 
&\sim \mathcal{N}(0, 2\rho^2) \\	
&\equiv \sqrt{2} \mathcal{N}(0, \rho^2)	\nonumber
\end{align}
such that the multiple values of $\delta$ could be expressed as 
\begin{align}
\begin{aligned}
\label{eq:delta_sum}
\delta_1 \pm  \delta_2 &\equiv \sqrt{2}\delta,\\
\text{with}\; \delta &\sim \mathcal{N}(0, \rho^2)		
\end{aligned}
\end{align}
from which point the previous method can be followed. The equivalence in \eqref{eq:delta_sum} occurs because $\delta_1,\delta_2$ are independent samples from a Gaussian distribution, meaning their combination is also Gaussian distributed.

%
\section{DCG constructions employed in this work}
\label{app:DCG}
%

\begin{table*}[t!]
\center
\renewcommand{\arraystretch}{2}
    \begin{tabular}{|>{\bfseries}c||c|c|c|c|}
 	\hline
	\textbf{Gate Construction}
	& ($\theta_1, \Omega_1, \phi_1$)
	& ($\theta_2, \Omega_2, \phi_2$) 		
	& ($\theta_3, \Omega_3, \phi_3$)
	& ($\theta_4, \Omega_4, \phi_4$)	    	
	\\ \hhline{|=||=|=|=|=|}		
	Primitive
	& ($\theta_t , \Omega, 0 $)
	& - & - & -
	\\ \hline
	CORPSE
	& ($2\pi + \theta_t/2 - k , \Omega, 0 $)
	& 	($2\pi - 2k , \Omega, \pi $)  		
	& ($\theta_t/2 - k , \Omega, 0 $)
	& -
	\\ \hline
	WAMF
	& ($\frac{X_0 + X_3}{4} , \Omega, 0 $)	
	& ($\frac{X_0 - X_3}{2} , \frac{X_0 - X_3}{X_0 + X_3} \Omega, 0 $)  
	& ($\frac{X_0 + X_3}{4} , \Omega, 0 $)	
	& -
	\\ \hline
	BB1	
	& ($\theta_t , \Omega, 0 $)	
	& ($\pi, \Omega, \phi_k $)			
	& ($2\pi  , \Omega, 3\phi_k $)
	& ($\pi  , \Omega, \phi_k $) 
	\\ \hline							
	\end{tabular}
\caption{Gate parameters required to construct a target rotation about the $x$-axis by angle $\theta_t$ using different pulse constructions. An additional $\pi/2$ shift in $\phi$ is required for rotations about the $y$-axis. Here, \mbox{$k = \arcsin{[\tfrac{\sin{[\theta_t/2]}}{2}]}$}, \mbox{$\phi_k = \arccos{[\tfrac{-\theta_t}{4\pi}]}$}, and for WAMF DCGs, the target rotations \mbox{$\theta_t = (\tfrac{\pi}{4}, \tfrac{\pi}{2}, \pi)$} have \mbox{$X_0 = (2\tfrac{1}{4}, 2\tfrac{1}{2}, 3)\pi$} and \mbox{$X_3 = (0.36, 0.64, 1)\pi$} determined explicitly.}
\label{tab:DCG_constructions}
\end{table*}

Three error suppressing DCGs are utilised in this work: CORPSE and WAMF gates, which suppress detuning errors, and BB1 gates,  which suppress over-rotation errors. For each of these constructions, the target angle $\theta_t = \pi,\, \pi/2$ gates are created as multi-segment pulses described by the segments' rotation angles $\theta_i$, phase angles $\phi_i$, and Rabi frequencies $\Omega_i$ normalized to the maximum frequency $\Omega$. The constructions of the different gates are shown in Table \ref{tab:DCG_constructions}. To ensure that the error suppressing aspects of the DCGs are maintained for all Clifford gates, the identity gate is implemented as a rotary spin echo by concatenating a $\pi$ rotation about the $x$-axis with its inverse $-\pi$ rotation. While this results in a net zero rotation, effectively identical to the simple wait time used for primitive $\Id$ gates, it makes the identity operation first-order insensitive to detuning errors during its implementation. The physical motivation here is that if a qubit is remaining idle at any point during a multi-qubit circuit, it may be preferable to continuously drive this type of rotary spin echo to ensure that it does not accumulate phase errors during its idle period.

%
\section{Influence of Quantum Projection Noise}
\label{app:QPN}

Quantum projection noise (QPN) describes the intrinsic uncertainty in qubit measurements due to the binomial nature of measurement outcomes~\cite{Itano:1993} and its scaling with the number of samples. The variance of a measurement due to QPN is $\nicefrac{p(1-p)}{r}$, where $p$ is the true state projection onto the $z$-axis of the Bloch sphere and $r$ is the number of identical measurements performed. Our work studies variances over distributions of noise-averaged survival probabilities, and consequently it is necessary to demonstrate that we were not limited by QPN bounds.

\begin{figure}[b!]
	\centering
	\includegraphics[scale=1]{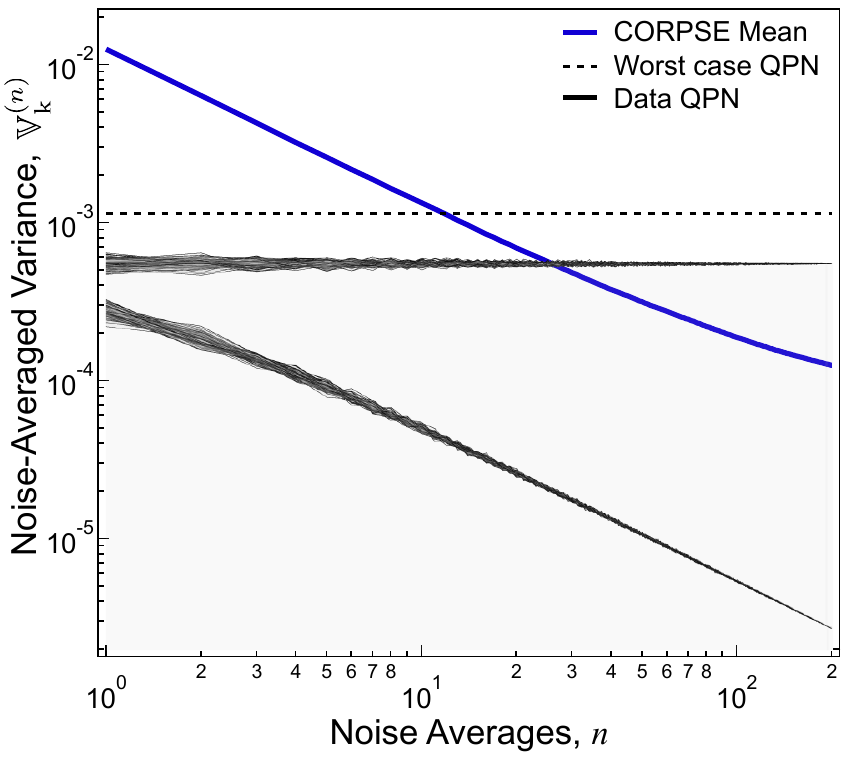}
	\caption{\textbf{Quantum projection noise limits for measured survival probabilities with the CORPSE DCG.}
Comparison of mean CORPSE variance scaling from main text Fig.~\ref{fig:DCG_engineerednoise}c (blue) to QPN variance limits given by $p(1-p)/r$. Dashed line is worst case QPN for $r=220$ when $p = 0.5$. Black lines show additional QPN limits where, for each $n$, $p(1-p)/r$ is calculated for 100 randomizations of noise realizations. The lower line scaling as $1/n$ is divided by $(n\times r)$ rather than $r$.}
	\label{fig:QPN}
\end{figure}

We consider the CORPSE data shown in Fig.~\ref{fig:DCG_engineerednoise}c; in order to ensure that our results are not measurement artefacts from quantum projection noise, we average each sequence and noise realization combination $r=220$ times. At this number of repetitions, the largest possible projection noise variance is given by $\nicefrac{0.5(1-0.5)}{220} = 1\times10^{-3}$. In addition to the worst case QPN, we compare the variance scaling results for the CORPSE DCG under simultaneously applied correlated and uncorrelated noise to the QPN given by the measured survival probabilities. Fig.~\ref{fig:QPN} shows the mean trajectory for the CORPSE variance scaling under the combined noise process presented in main text Fig.~\ref{fig:DCG_engineerednoise}c in dark blue. The dashed black line gives the worst case QPN and the two other sets of trajectories are calculated directly from the measured probabilities. For these, the QPN was calculated at each $n$ for 100 randomizations of noise realizations to reduce bias, and the 100 values are plotted. The lower set of trajectories are divided by $(n\times r)$ rather than just $r$. Our results are well above this lower limit suggesting that this is the most valid measurement of setting our QPN limit. Furthermore, we note that the saturation observed at large values of $n$ is not set by any static QPN bound limiting our measurements.

\end{document}